\documentclass[reprint,amsmath,amssymb,aps,superscriptaddress,twocolumn,10pt]{revtex4-1}
\usepackage[colorlinks=true,allcolors=blue]{hyperref}
\usepackage{graphicx,color}
\usepackage{dcolumn}
\usepackage{subfigure}
\usepackage{bm}
\usepackage{amsmath,amsfonts,amssymb}
\usepackage{acronym}
\usepackage{enumitem}
\usepackage{array}
\usepackage{multirow}
\usepackage{amsfonts}
\usepackage{amssymb}
\usepackage{amsthm,amsmath}
\usepackage{mathrsfs}
\usepackage{indentfirst}
\usepackage{multirow}
\usepackage{color}
\usepackage{ulem}

\allowdisplaybreaks

\newcommand{\TRC}{MOE Key Labortory of TianQin Mission,
TianQin Research Center for Gravitational Physics $\&$ School of Physics and Astronomy,
Frontiers Science Center for TianQin,CNSA Research Center for Gravitational Waves,
Sun Yat-sen University (Zhuhai Campus), Zhuhai 519082, China}

\newcommand{\PKUPHYS}{Department of Physics, Peking University, No.5 Yiheyuan Road, Beijing 100871, People's Republic of
China}

\newcommand{\PurpleObservatory}{
Key Laboratory of Dark Matter and Space Astronomy, Purple Mountain Observatory, Chinese Academy of Sciences, Nanjing 210023, People's Republic of China}

\newcommand{\SOUTHCUT}{
School of Physics and Optoelectronics, South China University of Technology, Guangzhou 510641,
People's Republic of China}

\newcommand{\PKUHIGHPHYS}{
Center for High Energy Physics, Peking University,
No.5 Yiheyuan Road, Beijing 100871, People's Republic of China}

\newcommand{\QuantumMater}{
Collaborative Innovation Center of Quantum Matter,
No.5 Yiheyuan Road, Beijing 100871, People's Republic of China}

\newacro{EMRI}{Extreme Mass Ratio Inspiral}
\newacro{MBH}{massive black hole}
\newacro{MCO}{massive compact object}
\newacro{MECO}{massive exotic compact object}
\newacro{DWD}{double white dwarf}
\newacro{AK}{analytic kludge}
\newacro{NK}{numerical kludge}
\newacro{AAK}{augmented analytic kludge}
\newacro{CO}{compact object}
\newacro{SNR}{signal-to-noise ratio}
\newacro{PN}{post newtonion}
\newacro{FIM}{Fisher information matrix}
\newacro{LSO}{last stable orbit}
\newacro{GW}{gravitational wave}
\newacro{BBH}{Binary Black Hole}
\newacro{BNS}{Binary Neutron Star}
\newacro{NS}{Neutron Star}
\newacro{ISCO}{inner stable circle orbit}
\newacro{SNR}{signal-to-noise ratio}
\newacro{GR}{general relativity}

\def\be{\begin{equation}}
\def\ee{\end{equation}}

\def\bea{\begin{eqnarray}}
\def\eea{\end{eqnarray}}

\begin{document}
\title{
\textbf{Analytic Kludge Waveforms for Extreme Mass Ratio Inspirals of Charged Object around Kerr-Newman Black Hole}}

\author{Tieguang Zi}
\affiliation{\SOUTHCUT}
\affiliation{\TRC}
\author{Ziqi Zhou}
\affiliation{\PKUPHYS}
\author{Hai-Tian Wang}
\affiliation{\PurpleObservatory}
\affiliation{\TRC}
\author{Peng-Cheng Li}
\email{pchli2021@scut.edu.cn}
\affiliation{\SOUTHCUT}
\affiliation{\PKUHIGHPHYS}
\affiliation{\TRC}
\author{Jian-dong Zhang}
\email{zhangjd9@mail.sysu.edu.cn}
\affiliation{\TRC}
\author{Bin Chen}
\affiliation{\PKUPHYS}
\affiliation{\PKUHIGHPHYS}
\affiliation{\QuantumMater}

\begin{abstract}
We derive the approximate, ``analytic-kludge'' (AK) waveforms for the inspiral of a charged  stellar-mass compact object (CO) into a charged massive Kerr-Newman (KN) black hole (BH). The modifications of the inspiral orbit due to the charges in this system can be attributed to three sources: the electric force between the CO and the MBH, the energy flow of the dipole electromagnetic radiation, and the deformation of the metric caused by the charge of the MBH. All these are encoded explicitly in the fundamental frequencies of the orbits, which are calculated analytically in the weak-field regime.
By calculating  the mismatch between the waveforms for charged and neutral EMRI systems with respect to  space-borne detectors TianQin and LISA, we show that tiny charges in the system can produce distinct imprints on the waveforms.
Finally, we perform parameter estimation for the charges using the Fisher information matrix method and find that the precision can reach the level of $10^{-5}$ in suitable scenarios. We also study the effects of charges on the parameter estimation of charge, where the effects from the charge of the MBH can be well explained by its effects on the cutoff of the inspiral.
\end{abstract}

\maketitle

%\newpage
%\baselineskip 18pt
\section{Introduction}
Black hole (BH)  no-hair theorems \cite{Bekenstein:1996pn,Chrusciel:2012jk} imply that the astrophysical BHs in electrovacuum  are described by the Kerr-Newman (KN) metric \cite{Newman:1965my}, which can be uniquely  characterized by the mass, spin, and electric charge. It's widely believed that the astrophysical BHs have negligible electric charge, due to the neutralization by surrounding plasmas, quantum discharge effects, or electron-positron pair production \cite{Gibbons:1975kk,Blandford:1977ds}. However, unequivocally observational evidence for the neutrality of both stellar-mass and massive  BHs are still lacking. Besides, some novel mechanisms have been
proposed such that BHs could retain a large amount of charge. For example, relying on
the well known Wald mechanism \cite{Wald:1974np} by which a spinning BH immersed in an external magnetic
field acquires a stable net charge, it was shown in \cite{Levin:2018mzg} that a strongly-magnetized neutron star in such a binary
system will give rise to a large enough charge in the BH to allow for potentially observable
effects. Additionally, the charge parameter in the KN metric can be regarded as the other types of charge, including magnetic charge (via duality transformation) \cite{Preskill:1984gd,Liu:2020vsy}, a vector charge in  the scalar-tensor-vector gravity (also known as ``MOG'')\cite{Moffat:2005si,Moffat:2016gkd}, a hidden electromagnetic charge in the mini-charged dark matter model~\cite{Cardoso:2016olt}. In all these cases, the BHs can be charged. Thus the validation of tiny or null charge in black hole is essential for us to understand various problems in gravity and astrophysics, ranging from the third hair on black hole to the formation mechanism of real astrophysical BHs.

In contrast to the electromagnetic observations \cite{Zakharov:2014lqa,Zajacek:2018ycb}, gravitational wave (GW) observations offer a more robust way to place constraints on the charges, in the sense that the assumptions for the models used in the former usually contain more uncertainties. If BHs are indeed charged and described by the KN metric, the gravitational wave signatures from binary BHs will be modified. Recently, several works \cite{Wang:2020fra,Bozzola:2020mjx,Bozzola:2021elc,Wang:2021uuh,Li:2021zct,Gupta:2021rod,Carullo:2021oxn} have emerged to analyze the charges of the stellar mass BHs observed by LIGO and Virgo \cite{LIGOScientific:2018mvr,LIGOScientific:2020ibl}.
For example, by performing numerical relativity simulations of the coalescence of the charged binary BHs, the authors in Refs. \cite{Bozzola:2021elc,Bozzola:2020mjx} showed that GW150914  is compatible with having a charge-to-mass ratio smaller than $0.3$. With the accurate quasinormal mode spectrum obtained from solving numerically the coupled perturbation equations of the KN BHs \cite{Dias:2021yju}, Ref. \cite{Carullo:2021oxn} analyzed the ringdown signal of GW150914 and found that the charge-to-mass ratio is smaller than $0.33$, by restricting the mass and spin to values compatible with the analysis of the full signal. For the future space-based GW detectors, Ref. \cite{Bao:2019kgt} analysed the capability of constraining the STVG parameter $\alpha$ with the ringdown of MBHBs, and the corresponding constraint for the charge-to-mass ratio is about $10^{-2}$.

Instead of the  binary BHs with comparable mass, in this work we focus on the charge effect on the GW signals generated from the inspirals
of stellar-mass compact objects (COs) into massive black holes (MBH) in galactic nuclei, i.e., the extreme-mass-ratio
inspirals (EMRI) \cite{Amaro-Seoane:2012lgq,Amaro-Seoane:2020zbo} and study the detectability of  the charges of both the CO and BHs for future space-based GW detectors such as TianQin (TQ) \cite{TianQin:2020hid} and LISA \cite{LISA:2017pwj}. EMRI is one of the most important sources for future space-based GW detectors, as the waveform contains a wealth of information about the spacetime geometry of BHs and the parameters of the system can be measured very accurately \cite{Babak:2017tow,Berry:2019wgg,Fan:2020zhy,Zi:2021pdp}. The data analysis  requires the construction of accurate waveforms  to enable accurate extraction of EMRI parameters from
a signal.
%The detection of such GWs requires the construction of accurate waveforms to extract signals from noisy data via matched filtering.

Since the waveforms are  slow-footed to calculate in the relativistic region for EMRIs, it is full of the substantial challenges to model EMRI dynamics precisely
and constrain source parameters \cite{Pound:2021qin}.
Most parameter estimations on EMRI sources adopt  the ``kludge" waveforms \cite{Barack:2003fp,Babak:2006uv,Chua:2017ujo} to generate fast waveforms.
In the realm of modeling EMRI, the kludge is in a way the approximative and built-up model that adopts several sets of post-Newtonian formulas to produce waveforms efficiently. The kludge waveforms can reflect the main feature of accurate EMRI waveforms including some relativistic effects, such as orbital eccentricity and relativistic precession. Recently, the full relativistic EMRI waveforms have been developed in \cite{Chua:2020stf,Katz:2021yft}, which combine the speed of EMRI ``kludge'' models and the accuracy of the first order gravitational self-force models.

In the past decades, EMRI waveforms have been worked out in the alternative theories of gravity
\cite{Barausse:2006vt, Yunes:2011aa,Gair:2011ym,Pani:2011xj,Sopuerta:2009iy,Canizares:2012is,Moore:2017lxy,Liu:2020ghq} to test their feasibility in the strong field region. Recently, Ref. \cite{Zhang:2022hbt} calculated the EMRI waveform for a charged object inspiraling around a Schwarzschild BH in a circular orbit using the Teukolsky method. In this work, we would like to extend the analytic kludge (AK) method to the charged case by considering the inspiral of a charged CO into a charged MBH. From the point of view of testing the Kerr hypothesis, the KN BH can be treated as a representative model deviating from  the Kerr BH, and its metric is analytically known and well behaved in the full range of the deviation parameter. In the AK model, the CO is moving on a Keplerian ellipse with the orbital parameters (semi-latus rectum and eccentricity ) are slowly evolving under the influence of radiation reaction, and the relativistic precession of the orbital plane and the
perihelion  are included. All the evolution equations are obtained under the PN approximations. Then the waveform is generated with the well-known Peter-Mathews formula \cite{Peters:1963ux,Peters:1964zz} in the quadrupole approximation. We will show that the introduce of charges into the EMRI will modify the precession via the fundamental frequencies of the orbits of the charged CO. Moreover, the effects of the charges on the radiation reaction are twofold: the direct modification to the loss of the energy and angular momentum due to gravitational radiation and the occurrence of electromagnetic radiation. Physically, the modifications due to the presence of the charges in this system can be attributed to the electric
force between the CO and the MBH, the dipole electromagnetic radiation, and the deformation of the metric caused by the charge
of the KN BH. The complete evolution equations of various orbital parameters are then obtained by combining the corrections due to the charges in the system appearing at the leading order of the PN expansion and the higher order terms from the original AK model. Furthermore, to
quantify the effects of the charges on the waveforms, we will compute the mismatch between  waveforms from  the neutral EMRI system and the charged one. Finally, we will perform parameter estimation of the charges for the CO and MBH with space-borne GW detectors
TQ and LISA.

The paper is organized as follows. In Sec. \ref{method}, we present the calculation of
the waveforms by following the approach of \cite{Barack:2003fp}. In Sec. \ref{signalana} we describe the formalisms of signal analysis for extracting values of the system parameters from the waveforms.
In Sec. \ref{results} we present the result of the constraints using the EMRI observation of LISA and TQ,
then place the parameter estimation about charges of the CO and the MBH.
Finally, we give a brief summary in Sec. \ref{summary}. Throughout this paper we use the geometric  units, where $c=G=1$.

\section{EMRI Waveforms}\label{method}
\subsection{Equations of motion}
The KN BH is a stationary, axisymmetrical, and asymptotically
flat solution of the Einstein-Maxwell equation. In Boyer-Lindquist coordinates,
the KN metric can be written as
\begin{eqnarray}
d s^2 &=&\frac{\Sigma}{\Delta} d r^2 + \Sigma d \theta^2 + \frac{\sin^2
  \theta}{\Sigma} [(r^2 + \tilde{a}^2) d \phi - \tilde{a} d t]^2
\\ &-& \frac{\Delta}{\Sigma} [\tilde{a}
  \sin^2 \theta d \phi - d t]^2,
\end{eqnarray}
where
\begin{eqnarray}
  \Sigma (r, \theta)  = r^2 + \tilde{a}^2 \cos^2 \theta, \\
  \Delta (r)  =  r^2 - 2 M r + \tilde{a}^2 + \tilde{Q}^2,
\end{eqnarray}
and $M$ and $\tilde{Q}$ are the mass and electric charge of the BH, and $\tilde{a}$ is specific angular momentum.
The electromagnetic potential is given by
\be A = \frac{\tilde{Q} r}{\Sigma} (d t - \tilde{a} \sin^2 \theta d \phi) .
\ee
The first-order equations of motion for  a timelike charged particle with mass $m$ and electric charge $\tilde{q}$  are given by \cite{Hackmann:2013pva}
\begin{eqnarray}
   \Sigma \frac{d t}{d \tau} & = & \frac{r^2 + M^2 a^2}{\Delta} P - a (a E \sin^2
  \theta - L_z)M^2, \\
  \Sigma \frac{d r}{d \tau} & = & \pm \sqrt{R}, \\
  \Sigma \frac{d \theta}{d \tau} & = & \pm M\sqrt{\Theta}, \\
  \Sigma \frac{d \phi}{d \tau} & = & \frac{a M}{\Delta} P - a M E +
  \frac{M L_z}{\sin^2 \theta},
\end{eqnarray}
where
\begin{eqnarray}
  R & = & P^2 - \Delta [r^2 + M^2(L_z - a E)^2 +M^2 C], \nonumber\\
  \Theta & = &C - \left[ (1- E^2) a^2 + \frac{L_z^2}{\sin^2 \theta}
  \right] \cos^2 \theta,
\end{eqnarray}
with
\begin{equation}
  P = E (r^2 + M^2 a^2) -M^2 a L_z -M q Q r.
\end{equation}
In the equations of motion there are three constants of motion. $\tilde{E}$ and $\tilde{L}_z$ are the conserved total energy and component of angular momentum parallel to symmetry axis, respectively, and $\tilde{C}$ is the  Carter constant \cite{Carter:1968rr}.
Note that in above expressions, we have used dimensionless quantities defined by
%\be \label{dimensionless}
\begin{align}\label{dimensionless}
 E= \frac{\tilde{E}}{m}, \quad a = \frac{\tilde{a}}{M}, \quad  q=\frac{ \tilde{q}} { m},
 \quad   L_z = \frac{\tilde{L}_z}{m M}, \notag\\
 C = \frac{\tilde{C}}{m^2 M^2}, \quad
 Q= \frac{\tilde{Q}}{M}.
\end{align}
One can see that the distinction between the above equations and the geodesic equations for a neutral particle moving in Kerr spacetime only reflects in the functions $\Delta$ and $P$.

%  \notag\\
\subsection{Fundamental frequencies }
In the AK model, if the spin of the CO is neglected, an EMRI event is completely specified by
$14$ degrees of freedom. To obtain the inspiral orbits of the charged CO in the KN background,
we append two charge  parameter $Q$ and $q$ to the  original AK model parameter space, such that
\begin{align}\label{parameters}
\lambda^i \equiv~ &\Big[\lambda^1,\cdots, \lambda^{16}\Big]   \notag\\
\quad  =\Big[&m, M, a,q ,Q ,e_{\rm LSO}, \tilde{\gamma}_0,\Phi,\cos\theta_S, \phi_S,\cos\lambda_I,\alpha_0, \nonumber\\
 &\cos\theta_K,\phi_K, D,t_0\Big],
\end{align}
where the definition and meaning  of each parameter can be found in \cite{Barack:2003fp}.

The trajectories of charged CO are roughly treated as quasi-Keplerian ellipses, which are characterized  by  the eccentricity $e$, and the radial orbital frequency $\nu$. The instantaneous phase of the CO in the orbit is specified by
the mean anomaly $\Phi$. The orientation of the orbit is described by three angles, $\lambda_I$, the inclination angle of the orbital plane with respect to the BH's spin direction $\hat{S}$, $\tilde{\gamma}$, the angle from pericenter  to $\hat{L}\times \hat{S}$  and $\alpha$ describing the direction of $\hat{L}$ around $\hat{S}$, where $\hat{S}$ is a unit vector of BH's spin and $\hat{L}$ is a unit vector of the orbital angular momentum.  The rate of change $\dot{\Phi}$, the orbital plane precession (also known as Lense-Thirring precession) frequency
$\dot{\alpha}$ and  the angular rate $\dot{\alpha}+\dot{\tilde{\gamma}}$
 of  periapsis precession are closely related to the fundamental frequencies of the CO's orbit by
\begin{eqnarray}
 \dot{\Phi} &=& 2 \pi \nu  =  \Omega_r, \\
  \dot{\tilde{\gamma}} & = & \Omega_{\theta} - \Omega_r, \\
  \dot{\alpha} & = & \Omega_{\phi} - \Omega_{\theta},
\end{eqnarray}
where $\Omega_r$, $\Omega_{\theta}$ and $\Omega_{\phi}$ denote the fundamental frequencies of radial, polar and azimuthal motion, respectively.
The closed form of these fundamental frequencies for Kerr BH orbits was first obtained by Schmidt \cite{Schmidt:2002qk} by employing the  elegant action-angle variable formalism of the Hamilton-Jacobi theory. Later on, combining
Schmidt's description and  using the Mino time \cite{Mino:2003yg}, Drasco and Hughes \cite{Drasco:2003ky} derived the
fundamental frequencies and showed the construction of the frequency domain representation of arbitrary functions
of orbits. In this work we would like to follow the steps in \cite{Drasco:2003ky} and also \cite{Fujita:2009bp,Gair:2011ym} to derive the
analytical expressions of these three fundamental frequencies in the weak-field regime.

First of all, in terms of the dimensionless time variable, i.e. the so-called Mino's time \cite{Mino:2003yg}
\be\label{minotime}
\frac{d}{d\lambda}=\frac{\Sigma}{M}\frac{d}{d\tau},
\ee
the equations of motion for the charged CO now become
\begin{eqnarray}
   \frac{d r}{d \lambda} & = & \frac{\pm \sqrt{R}}{M}, \label{GeoRadial}\\
  \frac{d \theta}{d \lambda} & = & \pm \sqrt{\Theta}, \\
  \frac{d t}{d \lambda} & = & T_r+T_\theta , \label{tRadial}\\
 \frac{d \phi}{d \lambda} & = &\Phi_r+\Phi_\theta, \label{phiRadial}
\end{eqnarray}
where
\be
T_r=\frac{r^2 + M^2 a^2}{M \Delta} P ,\quad T_\theta=- a (a E \sin^2
  \theta - L_z)M
\ee
\be
\Phi_r= \frac{a }{\Delta} P - a  E ,\quad \Phi_\theta=\frac{ L_z}{\sin^2 \theta}.
\ee
One can see that the $r$ and $\theta$ motions are now apparently decoupled. Next, we parameterize the orbit in terms of the (dimensionless)
semi-latus rectum $p$, the eccentricity $e$ and a phase angle $\psi$ via
\be\label{rMp}
r=\frac{M p}{1+e\cos\psi},
\ee
where $\psi$ varies from $0$ to $2\pi$ as $r$ goes through a complete cycle. The two turning points of the radial motion,
\be
r_a=\frac{M p}{1+e},\quad r_p=\frac{M p}{1-e},
\ee
are apoapsis and periapsis of the elliptic orbits, respectively. Moreover, the third orbital parameter is the turning point of the polar motion,
$\theta_{tp}$, which is also called the inclination angle, since this is equivalent to $\lambda_I$ \cite{Gair:2011ym} \footnote{Another usual orbital inclination angle is defined in terms of the Carter constant by $\cos\lambda_I=\frac{L_z}{\sqrt{C+L_z^2}}$, which in the limit $a\to0$ becomes $\cos\lambda_I=L_z/L$ and one has $\theta_{tp}+(\mathrm{sgn}\,L_z)\lambda_I=\pi/2$, where $L$ is the total orbital angular momentum.}. It is useful to express the constants of motion as functions of these three orbital parameters. This is implemented by solving the three equations
\be
R(r_a)=R(r_p)=\Theta(\cos\theta_{tp})=0.
\ee
The asymptotic form of these constants of motion in the weak-field regime are given by
\begin{eqnarray}
  E& =& 1 + \frac{1}{2 p} (e^2 - 1) (1 - q Q)+ \mathcal{O}(p^{-2}), \\ \label{energy}
  L_z &=& \sqrt{p}  \sqrt{1 - q Q} \sin \theta_{tp}+ \mathcal{O}(p^{-1/2}), \\ \label{angularmomentum}
  C&=& p (1-qQ)\cos^2\theta_{tp}+ \mathcal{O}(p^0).
\end{eqnarray}
Note that the terms at higher order in the $1/p$ expansion are not explicitly shown here, but they are important in the following calculations.

For bound orbits, $r(\lambda)$ and $\theta(\lambda)$ become periodic functions. Then from eq. (\ref{GeoRadial}), the fundamental period for the radial motion with respect to $\lambda$ is given by
\be
\Lambda_r=\int_0^{\Lambda_r} d\lambda=2\int_{r_a}^{r_p}\frac{M dr}{\sqrt{R}}=\int_0^{2\pi}\frac{d\psi}{\sqrt{V_\psi}},
\ee
where we have transformed the variable of the integral from $r$ to $\psi$, as the integral is easier to perform with the latter. The potential $V_\psi$ can be obtained through $R$ and (\ref{rMp}).

Similarly, for the polar motion, the fundamental period is given by
\be
\Lambda_\theta=4\int_{\theta_{tp}}^{\pi/2}\frac{d\theta}{\sqrt{\Theta}}=\int_0^{2\pi}\frac{d\chi}{\sqrt{V_\chi}},
\ee
where we have introduced the variable $\chi$ via $\cos^2\theta=\cos^2\theta_{tp}\cos^2\chi$, such that
as $\chi$ varies from $0$ to $2\pi$, $\theta$ oscillates through its full range of motion, from $\theta_{tp}$ to $\pi-\theta_{tp}$ and back \cite{Drasco:2003ky}. Thus, the angular
frequencies of the radial and the polar motion with respect to $\lambda$ then become
\be
\omega_r=\frac{2\pi}{\Lambda_r},\quad \omega_\theta=\frac{2\pi}{\Lambda_\theta}.
\ee
For the azimuthal motion, the equation (\ref{phiRadial}) is the sum of a function of $r$ and
a function of $\theta$, which allows us to define the frequencies of the coordinate $\phi$ with respect to $\lambda$ as
\be
\omega_\phi=\left<\frac{d\phi}{d\lambda}\right>_\lambda=\left<\Phi_r\right>_\lambda+\left<\Phi_\theta\right>_\lambda,
\ee
where
\be
\left<\Phi_r\right>_\lambda=\frac{1}{\Lambda_r}\int\Phi_rd\lambda=\frac{1}{\Lambda_r}\int_0^{2\pi}\frac{\Phi_r}{\sqrt{V_\psi}}d\psi,
\ee
and
\be
\left<\Phi_\theta\right>_\lambda=\frac{1}{\Lambda_\theta}\int_0^{2\pi}\frac{\Phi_\theta}{\sqrt{V_\chi}}d\chi.
\ee
Analogously, for the motion in $t$, the  equation (\ref{tRadial}) is also the sum of a function of $r$ and
a function of $\theta$, so we can define the frequencies of the coordinate $t$ with respect to $\lambda$
as
\be
\omega_t=\frac{1}{\Lambda_r}\int_0^{2\pi}\frac{T_r}{\sqrt{V_\psi}}d\psi+\frac{1}{\Lambda_\theta}\int_0^{2\pi}\frac{T_\theta}{\sqrt{V_\chi}}d\chi.
\ee
Up to now, the above fundamental frequencies were written with respect to the Mino time $\lambda$, the frequencies with respect to the distant observer time, i.e. the coordinate time $t$, are obtained by \cite{Schmidt:2002qk,Drasco:2003ky}
\be
\Omega_r=\frac{\omega_r}{\omega_t},\quad \Omega_\theta=\frac{\omega_\theta}{\omega_t},\quad \Omega_\phi=\frac{\omega_\phi}{\omega_t}.
\ee
Explicitly,  the asymptotic form of these fundamental frequencies in the weak-field regime are given by
\begin{eqnarray}
\Omega_r & = &\frac{\sqrt{1 - q Q}}{M} \left( \frac{1 - e^2}{p} \right)^{3/2} +
 \frac{3 \sqrt{1 - q Q} (q Q - 4)}{4 M}  \left( \frac{1 - e^2}{p}  \right)^{5 / 2}  \nonumber\\
  &+& \mathcal{O} (p^{- 3}), \\
\Omega_{\theta}  &=& \frac{\sqrt{1 - q Q}}{M} \left( \frac{1 - e^2}{p}\right)^{3/2}
    + \frac{a (1 - q Q) (e^2 - 1)^3 \sin \theta_{tp}}{M p^3} \nonumber\\
     &+& \frac{1}{M} \frac{\left.Q \left(q^2 Q-3 q+2 Q\right)-3 e^2 \left(q^2 Q^2-5 q Q+4\right)\right)}{4 (1 - e^2)
  \sqrt{1 - q Q}}\left( \frac{1 - e^2}{p} \right)^{5/2}
 \nonumber\\ &+&  \mathcal{O} (p^{- 7/2}), \\
\Omega_{\phi} &=& \frac{\sqrt{1 - q Q}}{M} \left( \frac{1 - e^2}{p} \right)^{3/2}
\nonumber\\&+& \frac{a (2 - q Q) (1 - e^2)^{3 / 2} + a (1 - q Q) (e^2 - 1)^3 \sin\theta_{tp}}{M p^3}
\nonumber\\ & +& \frac{-Q (-3 q + 2 Q + q^2 Q) + 3 e^2 (4 - 5 q Q + q^2 Q^2)}{4M (1 - e^2)\sqrt{1 - q Q}}
 \left( \frac{1 - e^2}{p} \right)^{5 / 2}
\nonumber\\  &  +&\mathcal{O} (p^{- 7 / 2}) .
\end{eqnarray}
Therefore, we obtain the
perihelion precession frequency and the orbital plane
precession frequency at the leading order of the $1/p$ expansion
\begin{eqnarray}
  \dot{\tilde{\gamma}} & = & \frac{1}{M} \frac{6 - 6 Q q - Q^2 + q^2 Q^2}{2 (1
  - e^2) \sqrt{1 - q Q}} \left( \frac{1 - e^2}{p} \right)^{5 / 2},\label{gammat} \\
  \dot{\alpha} & = & \frac{a (2 - q Q) (1 - e^2)^{3 / 2}}{M p^3}\label{alphat} .
\end{eqnarray}
Comparing with the uncharged equations, the modification due to the charges in the system can be summarized as the electric force between the CO and the KN BH in the form $qQ$ and the deformation of the metric from the charge of the KN BH in the form $Q^2$. As we will see in the next subsection, the other contribution of the charges to the waveforms stems from the dipole electromagnetic radiation in the form $(Q-q)^2$.
\subsection{Fluxes}
Besides the fundamental frequencies, the other piece of the AK waveform is the change rates of the eccentricity $e$ and the radial orbital frequency $\nu$ with respect to the coordinate time, which are related to the energy flux and the angular momentum flux due to the gravitational
radiation and electromagnetic radiation.

For the gravitational radiation, the standard quadrupole formulas of the energy flux and the angular momentum flux were already derived by
Peters and Mathews \cite{Peters:1963ux,Peters:1964zz}
\be
\left<\frac{d E}{dt}\right>=\frac{1}{5\mu}\left<\frac{d^3Q_{ij}}{dt^3}\frac{d^3Q^{ij}}{dt^3}
-\frac{1}{3}\frac{d^3Q^i_i}{dt^3}\frac{d^3Q^j_j}{dt^3}\right>,
\ee
and
\be
\left<\frac{d L_i}{dt}\right>=\frac{2}{5\mu M}\epsilon_{ijk}\left<\frac{d^2Q_{jm}}{dt^2}\frac{d^3Q^{km}}{dt^3}\right>,
\ee
where $\mu$ is the reduced mass $\mu=m M/(m+M)\simeq m$, $E$ and $L_i$ are the dimensionless energy and angular momentum appeared in the equations of motion of the charged CO. $Q_{ij}$ is the familiar quadrupole moment tensor of mass (also called the inertia tensor)
\be
Q_{ij}=\mu x^i x^j,
\ee
where $x^i$ is the relative position vector  between the charged CO and the charged central BH, and in the weak-field regime one has
$x^i=(r \cos\phi\sin\theta, r\sin\phi\sin\theta,r\cos\phi)$. In addition, the angle-brackets mean the average over one cyclic motion in $r$, which via (\ref{rMp}), can be turned into the integral for $\psi$,
\be
\left<\frac{d E}{dt} \right>=\frac{1}{T}\int_0^{2\pi}\frac{dE}{dt}\frac{d\psi}{\dot{\psi}}, \quad T=\int_0^{2\pi}\frac{d\psi}{\dot{\psi}}.
\ee
In the following, without ambiguity we will just use $dE/dt$ and $dL_z/dt$ to denote the averaged ones. Then from above formulas and the equations of motion of the CO, we obtain the energy flux and angular
momentum flux loss of the charged particle due to the gravitational radiation
\begin{eqnarray}
\frac{d E}{d t} & = & -  \frac{32\eta}{5M p^5} (1-qQ)^3 (1 - e^2)^{3 / 2}
\nonumber\\& & \times
  \left( 1 + \frac{73}{24} e^2 + \frac{37}{96} e^4 \right), \\
\frac{d L_z}{d t} & = & - \frac{32\eta\sin\theta_{tp}}{5Mp^{7/2}} (1-qQ)^{5/2} (1 - e^2)^{3 / 2}
\nonumber\\& &  \times \left( 1 + \frac{7}{8} e^2 \right) ,
\end{eqnarray}
where the assumption that we have made is same to \cite{Barack:2003fp} in which $\theta_{tp}$ is constant at leading order in the $1/p$ expansion. To simplify the expressions we have introduced the symmetric mass ratio,
\begin{equation}
  \eta \equiv \frac{m M}{(m + M)^2} \simeq \frac{m}{M} .
\end{equation}
In terms of the relation between $p$ and the radial orbital frequency at leading order
\begin{equation}
  p = \frac{(1 - e^2)(1-qQ)^{1/3}}{(2 \pi M \nu)^{2 / 3}},
\end{equation}
the above two equations can be written as
\begin{eqnarray}
\frac{d E}{d t} & = & - \frac{32\eta}{5M}  (2 \pi M \nu)^{10 / 3} (1 - q
  Q)^{4/3} (1-e^2)^{-7/2}
 \nonumber\\ &&  \times
\left( 1 + \frac{73}{24} e^2 + \frac{37}{96} e^4 \right), \label{GradEt}\\
\frac{d L_z}{d t} & = & - \frac{32 \eta \sin\theta_{tp}}{5M} (2 \pi M \nu)^{7/3} (1 -
  q Q)^{4 / 3} (1-e^2)^{- 2}
 \nonumber\\ &&  \times\left( 1 + \frac{7}{8} e^2 \right) .\label{GradLzt}
\end{eqnarray}
On the other hand, from the asymptotic form of the energy and angular momentum in the weak-field regime
(\ref{energy}) and (\ref{angularmomentum}),
we can obtain their change rates with respect to distant observer time
%\begin{widetext}
\begin{eqnarray}
  \frac{d E}{d t} & = & - \frac{(2 \pi M)^{2 / 3} (1 - q Q)^{2/3}}{3 \nu^{1 / 3}}
  \frac{d \nu}{d t}, \\ \label{GradEt:dot}
\frac{d L_z}{d t} & = & - \frac{(1 - q Q)^{2/3}\sin \theta_{tp}}{3(2 \pi M \nu)^{1/3} \sqrt{1 - e^2 } \nu}
\nonumber\\ &&
\times \left( 3e\nu\frac{d e}{d t} +(1-e^2) \frac{d \nu}{d t} \right) .\label{GradLzt:dot}
\end{eqnarray}
We should also take the orbit decay due to the electromagnetic radiation into account. The formulas of the energy and angular emission rates for the dipole electromagnetic radiation\footnote{Here we use the convention in \cite{Landau:1971},
where the Lagrangian of the electromagnetic field is
$\mathcal{L}=-\frac{1}{16\pi}F_{\mu\nu}F^{\mu\nu}+A_\mu J^
\mu$
and the equation of motion is $\nabla_\mu F^{\mu\nu}=-4\pi J^\mu$.} are given by \cite{Landau:1971} and also \cite{Liu:2020cds,Christiansen:2020pnv,Liu:2020bag,Liu:2022wtq}

\be
m\frac{dE_{EM}}{dt}=\frac{2}{3}\mu^2(Q-q)^2\left<\frac{d^2x^i}{dt^2}\frac{d^2x_i}{dt^2}\right>,
\ee
and
\be
m M\frac{dL_z^{EM}}{dt}=\frac{2}{3}\mu^2(Q-q)^2
\epsilon^{\jmath kl}\left<\frac{dx_k}{dt}\frac{d^2x_l}{dt^2}\right>,
\ee
where as before the angle-brackets denote the average over one period in $r$ motion, and $\jmath=z$.
It should be noted that the
above mentioned  $E^{EM}$ and $L_z^{EM}$ are set to be dimensionless, their relation with the original ones are the same as (\ref{dimensionless}).
Then for a charged particle orbiting a KN BH, the energy flux and angular
momentum flux loss due to the electromagnetic radiation at the leading order in $1/p$ expansion are given by
\begin{eqnarray}
  \frac{d E_{EM}}{d t} & = & - \frac{(2 + e^2) \eta^2}{3 (1 - e^2)^{5/2}} (2 \pi mM \nu)^{8 / 3}(Q - q)^2
\nonumber\\ && \times  (1 - q Q)^{2/3}, \label{EM:Edot}\\
  \frac{d L_z^{EM}}{d t} & = & - \frac{2 m  \sin\theta_{tp}}{3  M^3(1 - e^2)} (2\pi M \nu)^{5/3}(Q - q)^2
 \nonumber\\ &&\times   (1 - q Q)^{2 / 3}.\label{EM:Lzdot}
\end{eqnarray}
Combining the equations \eqref{GradEt}, \eqref{GradLzt}, \eqref{GradEt:dot}, \eqref{GradLzt:dot},
\eqref{EM:Edot} and \eqref{EM:Lzdot},
we have the evolution equations of orbital eccentricity $e$ and $\nu$  due to the gravitational and electromagnetic  radiation
\begin{eqnarray}
\frac{d \nu}{d t} &=& \frac{96}{10 \pi} \frac{\eta}{M^2} (2 \pi M \nu)^{11/3}(1-qQ)^{2/3}(1-e^2)^{-7/2}
 \nonumber\\ && \left(1+\frac{73}{24}e^2+\frac{37}{96}e^4\right)
\nonumber\\ && +\frac{\eta(2+e^2)(Q-q)^2(2 \pi M \nu)^3(1-e^2)^{-5/2}}{2\pi M^2},\\
\frac{de}{dt}&= &-\frac{e\eta}{15M}(2 \pi M \nu)^{8/3}(1-qQ)^{2/3}(1-e^2)^{-5/2}(304+121e^2)
\nonumber\\
&&-\frac{e\eta}{M}(Q-q)^2(2 \pi M \nu)^2(1-e^2)^{-3/2}
\end{eqnarray}
From the right hand sides of these equations, we find that the contribution from the dipole electromagnetic radiation is lower than that from the gravitational radiation by a factor of $(2 \pi M \nu)^{2/3}$, which for a Keplerian orbit corresponds to $v^2$, where $v$ is the orbital velocity of the CO. This means that the correction due to the dipole electromagnetic radiation appears at $-1$ PN order in the waveforms and becomes prominent at the early stage of the inspiral of the CO where $v$ is small. This is verified by numerical
relativity simulations of the coalescence of the charged binary BHs with comparable masses \cite{Bozzola:2020mjx}, where it was found that the greatest difference between charged and uncharged BHs arises in the earlier inspiral. In this case, despite the fact that the AK model is not accurate enough to produce EMRI template waveforms in the strong field region, the behavior of the electric charges may well be captured by this model.

In terms of the radial frequency $\nu$ the other two equations (\ref{gammat}) and (\ref{alphat}) are expressed as
\begin{eqnarray}
  \frac{d \tilde{\gamma}}{d t} & = & \frac{(6 - 6 Q q - Q^2 + q^2
  Q^2)}{(1 - q Q)^{4/3}(1 - e^2)} \pi \nu (2 \pi M \nu)^{2 / 3} , \\
  \frac{d \alpha}{d t} & = & \frac{ 4 a(2 - q Q) \pi^2 M \nu^2  }{(1 - e^2)^{ 3 / 2}(1-qQ)} .
\end{eqnarray}

From these four evolution equations we can see that the equations keep invariant under the operation $q\to-q$ and $Q\to-Q$, which means we cannot simultaneously determine the sign of $q$ and $Q$. Hereafter, without loss of generality,  we will always assume that the charge of the MBH is positive, and let the sign of the charge of the orbiting  particle to be free. For gravitational waveform generated by two charged compact objects  moving on a Keplerian orbit in a plane, the charges appear only in the form $(Q-q)^2$ and $qQ$, which hinders the unique determination of the two charges. However, the relativistic effects considered in the AK waveform breaks the degeneracy between these two terms. As a consequence, once the sign of $Q$ is provided, we can not only identify the value of $Q$ but also uniquely discern both the magnitude and the sign of $q$. Even the dipole electromagnetic radiation disappears when $Q=q$, this conclusion still works.

Up to now, we have obtained the leading order evolution equations for the relevant orbital parameters when both the CO and central BH are charged.
In analogy with the construction of EMRI waveforms  in alternative
theories of gravity, e.g. \cite{Zi:2021pdp}, we combine these leading order corrected equations with those higher-order PN equations in the original AK model.  Then the complete orbital evolution equations are given by
\begin{align}\label{evolution}
\dot{\Phi}~=~&2\pi \nu, \nonumber\\
\dot{\nu}~=~&\frac{48}{5\pi}\frac{\eta}{M^2}X^{11/3}Y^{-9/2} \Big\{(1-qQ)^{2/3}Y\Big(1+\frac{73}{24}e^2+\frac{37}{96}e^4\Big)
\nonumber\\&+X^{2/3}\Big[\frac{1273}{336}-\frac{2561}{336}e^2-\frac{3885}{128}e^4 -\frac{13147}{5376}e^6\Big] \nonumber\\& - X a \cos\lambda Y^{-1/2}\Big[\frac{73}{12}+\frac{1211}{24}e^2
+\frac{3143}{96}e^4+\frac{65}{64}e^6\Big] \Big\}
\nonumber\\&
+ \frac{\eta (Q-q)^2}{2\pi M^2}X^3(2+e^2)Y^{-5/2}
, \nonumber\\
\dot{e}~=~&-\frac{e \eta}{15M}Y^{-7/2}X^{8/3}\Big[((1-qQ)^{2/3}+12X^{2/3})(304+121e^2)Y
\nonumber\\&-\frac{1}{56}X^{2/3}(133640+108984e^2+25211e^4)\Big]
\nonumber\\& +e \frac{\eta}{M}a \cos\lambda X^{11/3}Y^{-4}\left(\frac{1364}{5}+\frac{5032}{15}e^2
+\frac{263}{10}e^4\right)
\nonumber\\& - e\frac{\eta}{M} (Q-q)^2 X^2 Y^{-3/2},\nonumber\\
\dot{\tilde{\gamma}}~ =~& \pi \nu X^{2/3}Y^{-1}\Big[(6-6qQ-Q^2+q^2Q^2)(1-qQ)^{-4/3}
\nonumber\\&
+\frac{3}{2}X^{2/3}Y^{-1}(26-15e^2)\Big]
 - 12\pi \nu a \cos\lambda X Y^{-3/2}
, \nonumber\\
\dot{\alpha}~=~& 4aM(\pi \nu)^2Y^{-3/2}(2-qQ)/(1-qQ)
\end{align}
where dot denotes the derivative with respect to time and to avoid redundant expression, we have defined $Y=1-e^2$, $X=2 \pi M \nu$. The equation for $\dot{\nu}$ and $\dot{e}$ are given accurately through $3.5$ PN order, the equations for $\dot{\tilde{\gamma}}$ and  $\dot{\alpha}$ are accurate through $2$ PN order.

We would like to give a brief review of how do above evolving orbital parameters enter the AK EMRI waveforms. The general GW strain field at the detector is written as
 \be
 h_{ij}(t)=A^+(t)H_{ij}^+(t)+A^\times(t) H^\times_{ij}(t),
 \ee
 where $H_{ij}^+$ and $H_{ij}^\times$ are the two polarization basis tensors constructed with the unit vector pointing from the detector to the source $\hat{n}$ and  the unit vector along the CO's
orbital angular momentum $\hat{L}$,
\be
H_{ij}^+(t)=\hat{p}_i\hat{p}_j-\hat{q}_i\hat{q}_j,\quad H_{ij}^\times(t)=\hat{p}_i\hat{q}_j+\hat{q}_i\hat{p}_j,
\ee
with
\be
\hat{p}=\frac{\hat{n}\times\hat{L}}{|\hat{n}\times\hat{L}|},\quad \hat{q}=\hat{p}\times\hat{n},
\ee
and $A^+$ and $A^\times$ are the amplitudes of the two polarizations. The amplitudes of the two polarisations can be further written in terms of the
Peters-Mathews harmonic decomposition as
\begin{align}
A^+  & \equiv  \sum_n A_n^+  \notag\\
 &= \sum_n -\Big[1 + (\hat{L} \cdot \hat{n})^2\Big]\Big[a_n \cos2\gamma -b_n\sin2\gamma\Big]
\nonumber\\   & +  c_n\Big[1-(\hat{L} \cdot \hat{n})^2\Big], \\
A^\times  &\equiv   \sum_n A_n^\times  \notag\\
& = \sum_n 2 (\hat{L} \cdot \hat{n}) \Big[b_n\cos2\gamma+a_n\sin2\gamma\Big],
\end{align}\label{amplitude}
where  $(a_n,b_n,c_n)$ come from decomposition of the second time derivative of the inertia tensor $Q^{ij}$ into $n-$harmonics of the radial orbital frequency $\nu$ and are functions of $\nu$ and $e$ \cite{Peters:1963ux}. Moreover, $\gamma$ is an azimuthal angle measuring the direction of pericentre with respect to
the orthogonal projection of $\hat{n}$ onto the orbital plane, which further depends on $\tilde{\gamma}$ and $\alpha$.

Since the equilateral triangle detectors such as TQ can be used to construct two independent Michelson
interferometers, the signal responded by such two interferometers can be written as:
\be
h_{I,II}=\frac{\sqrt{3}}{2}\left(F^+_{I,II}h^++F^\times_{I,II}h^\times\right),
\ee
where the antenna pattern function $F^{+,\times}_{I,II}$ of the detector depend on the orbits of satellites \cite{Cutler:1994ys}.
For TQ, the detailed information of the respond function for EMRI signal can be found in \cite{Fan:2020zhy}.

\begin{figure}[htbp]
	\centering
	\includegraphics[width=0.48\textwidth]{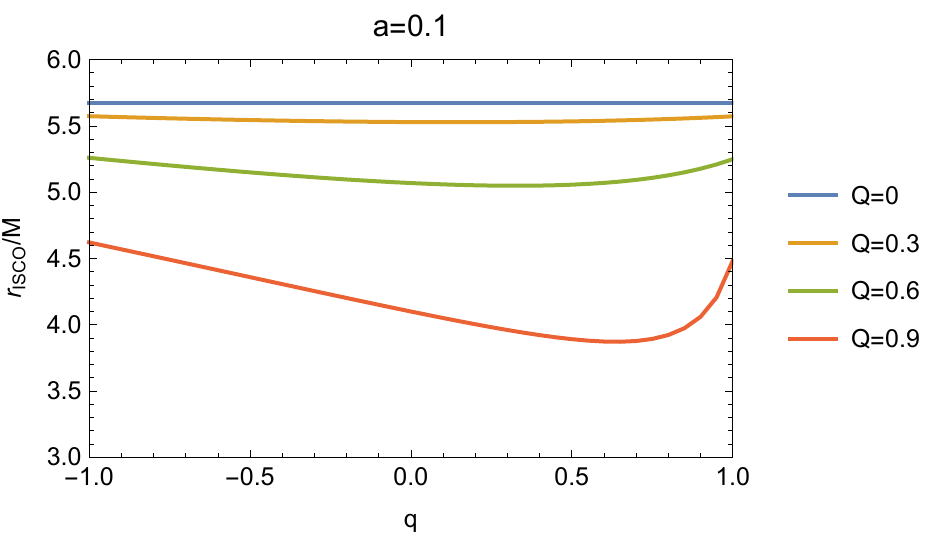}
	\includegraphics[width=0.48\textwidth]{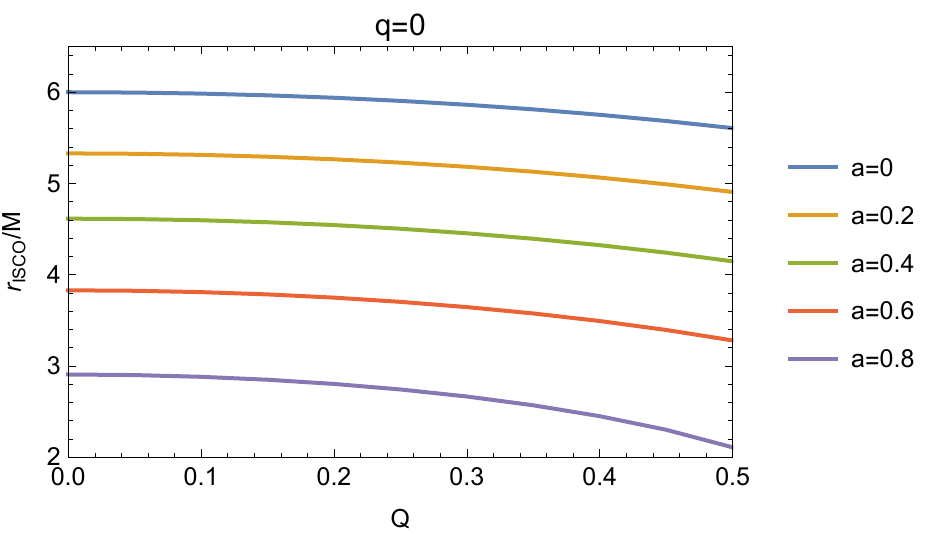}
	\caption{The influence on the radius of ISCO by the charge of CO and MBH, and the spin of MBH. Above: The $r_\text{ISCO}$ as a function of $q$ for different $Q$. Below: The $r_\text{ISCO}$ as a function of $Q$ for different $a$.}\label{ISCOMBH}
\end{figure}

At the final stage of EMRI, when the CO passed the
boundary of stable orbits, it will plunge into the MBH
directly in a short time. So we need to introduce a cutoff frequency to the waveform. When the CO is moving in the equatorial plane of the central BH, the cutoff is usually taken to be the last stable orbit (LSO). Here for simplicity we take the  innermost stable circular
orbit (ISCO) as the cutoff in our waveform. The radius of the ISCO results from the equations $R(r)=R'(r)=R''(r)=0$. Although these equations can be solved analytically with {\em Mathematica}, the expression of the solution is very lengthy so we shall not show it here. Instead, the effects of the  MBH charge $Q$ and CO charge $q$ on the ISCO radius can be demonstrated clearly in a graphical manner. Note that in this work we only consider the  prograde orbits of the CO, since most of
the detected events have prograde orbits \cite{Fan:2020zhy}.

As shown in Fig. \ref{ISCOMBH}, $r_{\rm ISCO}$ changes  gradually with the MBH charge $Q$  (left panel)
and CO charge $q$  (right panel). This indicates that the CO gets a chance to orbit more circles in the vicinity of the KN BH, and the charged EMRI system radiates higher frequency GW signal than the neutral system. The effect of the CO charge on $r_{\rm ISCO}$ is more complicated. In this case, $r_{\rm ISCO}$ no longer depends on $q$ monotonically. Nevertheless, the turning point at which the monotonicity of $q$ changes increase with $Q$.

\section{Method of GW analysis}\label{signalana}
In this section we present some recipes for the assessing of the impacts the charge parameters $q$ and $Q$ on the EMRI waveforms and the constraint on them  with LISA and TianQin observations of EMRIs.

%\subsection{Mismatch}
%\begin{figure*}[th]
%\centering
%\includegraphics[width=0.45\textwidth]{lisa-Mismatch-mbhelec.pdf}
%\includegraphics[width=0.45\textwidth]{lisa-Mismatch-COelec.pdf}
%\caption{Missmatch $\mathcal{M}$ as a function of MBH charge $Q$ (left panel) and CO charge $q$ (right panel) is plotted using %observation of LISA.}
%\label{mismatch:lisa}
%\end{figure*}
%

%\begin{figure*}[th]
%	\centering
%	\includegraphics[width=0.45\textwidth]{TQ-overlap-MBHelec.pdf}
%	\includegraphics[width=0.45\textwidth]{TQ-overlap-COelec.pdf}
%	\caption{Missmatch $\mathcal{M}$ as a function of MBH charge $Q$ (left panel) and CO charge $q$ (right panel) is plotted using observation of TQ.
%		}\label{mismatch:tq}
% \end{figure*}
To assess the effects the charges of the MBH and CO on the EMRI waveforms,
it is convenient to define the overlap $\mathcal{O}$ between two sets of waveforms $h_a(t)$ and $h_b(t)$,
\begin{equation}
\mathcal{O}(h_a|h_b)=\frac{<h_a|h_b>}{\sqrt{<h_a|h_a><h_b|h_b>}},
\end{equation}
where the inner product $<h_a|h_b>$ is defined by
\begin{align}\label{inner}
<h_a |h_b > =2\int^\infty_0 df \frac{h_a^*(f)h_b(f)+h_a(f)h_b^*(f)}{S_n(f)}
\end{align}

Here $h_a(f)$ is the Fourier transform of the time domain waveform $h_a(t)$, $*$ denotes complex conjugate  and $S_n(f)$
is noise power spectral density of space-borne GW detectors, such as LISA~\cite{LISA:2017pwj} and TianQin (TQ)~\cite{TianQin:2020hid}.
Hence we get the mismatch $\mathcal{M}$ for two different waveforms,
\begin{equation}
    \mathcal{M} \equiv 1- \mathcal{O}(h_a|h_b).
\end{equation}
Obviously, if the two waveforms are identical, the overlap between them equates to unity and so their mismatch is zero.  For a signal with signal-to-noise ratio (SNR) $\rho$, the
mismatch of two different waveforms has to be larger than $\mathcal{D}/2\rho^2$ for a detector to distinguish them \cite{Flanagan:1997kp,Lindblom:2008cm}, where $\mathcal{D}=9$ denotes the number of the intrinsic parameters of the EMRI system. The intrinsic parameters describe the system without reference to the location or orientation of
the observer \cite{Buonanno:2002ft}.
For example, the SNR threshold for EMRI that can be detected is usually chosen to be $20$ \cite{Babak:2017tow}. Then the mismatch of two waveforms larger than $0.01125$ can be resolved for a EMRI event which has just reached the threshold of detection.

\subsection{Fisher informational matrix}
To quantify the capability of space-borne GW detectors  to constrain the
 MBH and CO charges, we  use the fisher informational matrix (FIM) method \cite{Vallisneri:2007ev} to obtain the lowest-order expansion of the posteriors (valid in the high SNR limit), which can be more accurately estimated through a
full Bayesian parameter estimation analysis.

\begin{figure*}[th]
\centering
\includegraphics[width=0.45\textwidth]{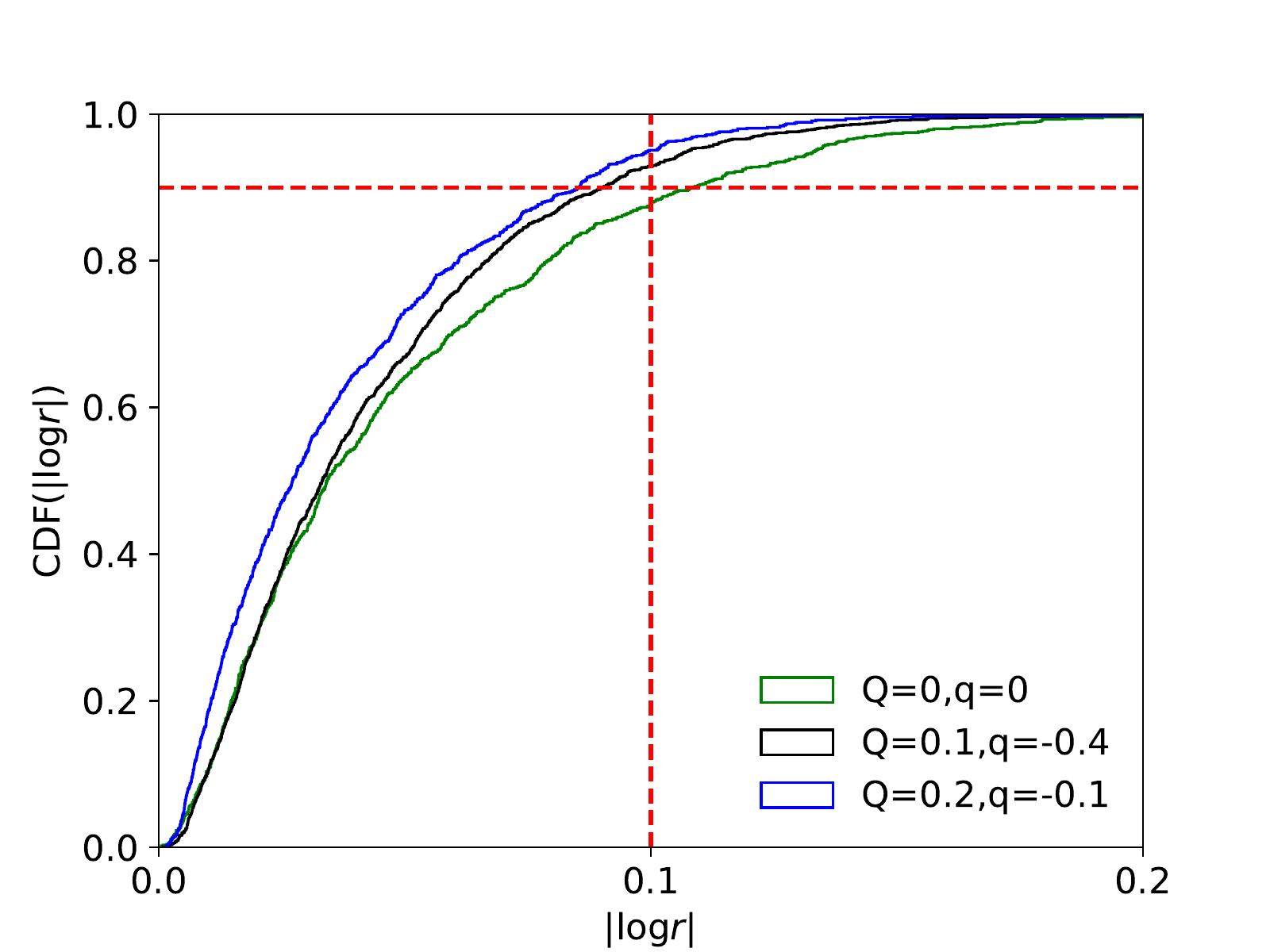}
\caption{ Cumulative distribution function of logarithm of mismatch ratio $\log(r)$ for different charges, assuming the SNR $\rho=20$, is plotted. The horizontal and vertical dashed line represents $90\%$ point of CDF of $|\log(r)|$ and $|\log(r)|=0.1$, respectively. The FIM is valid when the curve surpassed these dashed lines.  The other parameters take values as those in Sec.\ref{source:params}.  }\label{CDF}
\end{figure*}

The FIM is defined by
\begin{equation}
    \Gamma_{ab}=\Big( \frac{\partial h}{\partial \lambda_a} \Big| \frac{\partial h}{\partial \lambda_b} \Big),
\end{equation}
where $\lambda_a$, $a=1,2,...,$ are the parameters appearing in the waveform (\ref{parameters}) and  the inner product $(|)$ is  defined by equation \eqref{inner}.
When the SNR of the GW signal is large, the variance-covariance matrix can be
obtained as the inverse of the FIM
\be
\Sigma_{ab}\equiv <\Delta \lambda_a \Delta\lambda_b>=(\Gamma^{-1})_{ab}.
\ee
From the variance-covariance matrix, the uncertainty $\sigma_a$ of the $a$th parameter $\lambda_a$ can be
obtained as
\begin{eqnarray}
\delta\lambda_a=\Sigma_{aa}^{1/2}.
\end{eqnarray}
Note that the  applicability of the FIM method requires the  linear signal approximation to be valid \cite{Vallisneri:2007ev}, so strictly speaking, we should verify this point. By following the procedure in \cite{Vallisneri:2007ev} and for LISA, we calculate the cumulative distribution function for mismatch ratio $r$, which characterizes the difference between the actual value of likelihoods and the linear signal approximation. The criterion is  when $|\log r|<0.1$ over $90\%$ of the $1\sigma$ surface for a given SNR, we can say the FIM method is valid.
In Fig. \ref{CDF}, we show the cumulative distribution function (CDF)
for logarithm of $r$ at SNR $\rho=20$, with $Q=q=0$, $Q=0.1, q=-0.4$ and $Q=0.2,q=-0.1$.
We can see  that for most of the random points at the $1\sigma$ surface,
the derived value of likelihood using FIM slightly deviations from the exact likelihood, which means the parameter estimation for charged EMRI system with FIM method is basically valid.

\section{Results}\label{results}
\subsection{Waveforms and mismatch}\label{source:params}
\begin{figure*}[th]
\centering
    \includegraphics[width=0.45\textwidth]{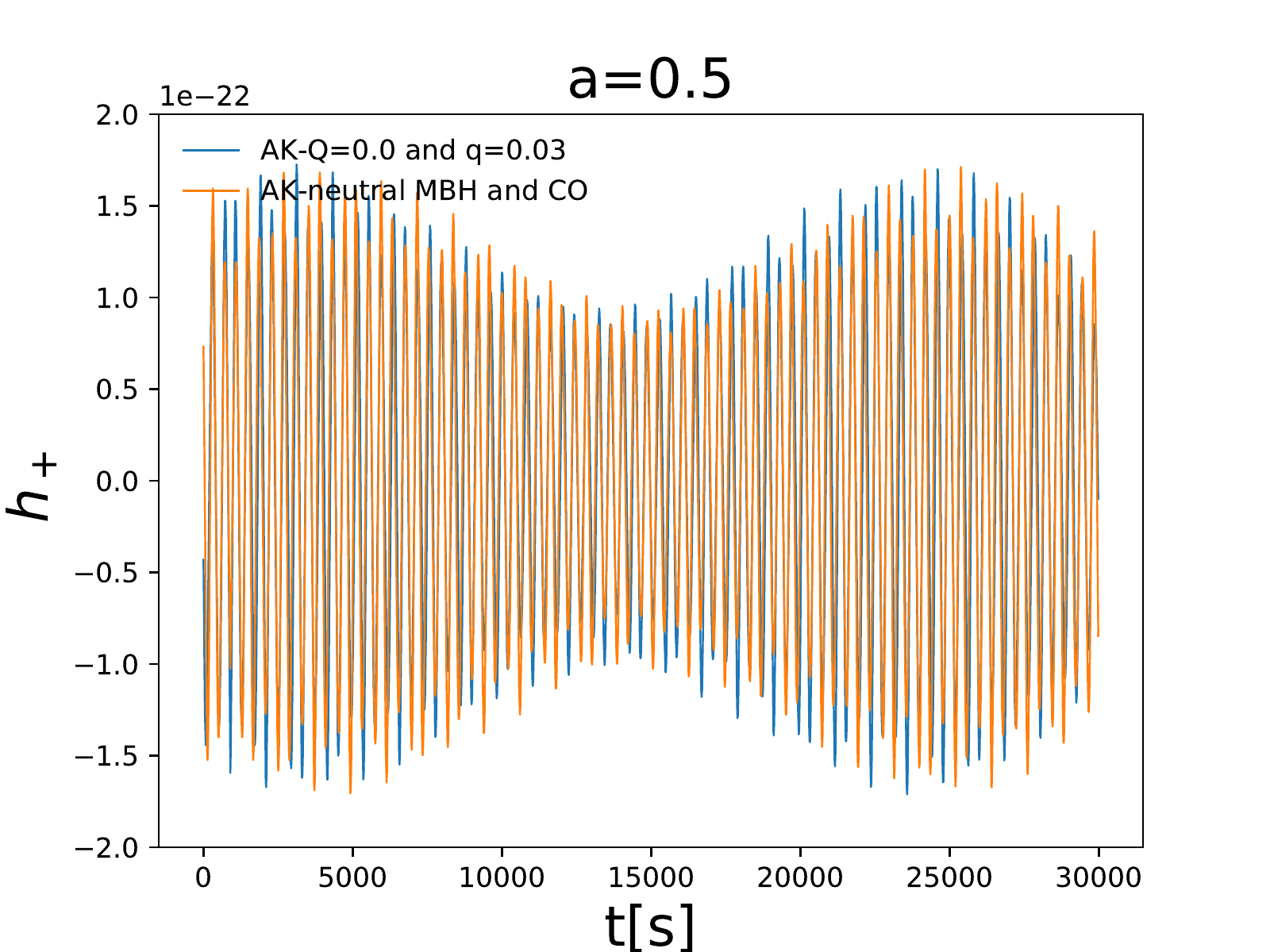}
	\includegraphics[width=0.45\textwidth]{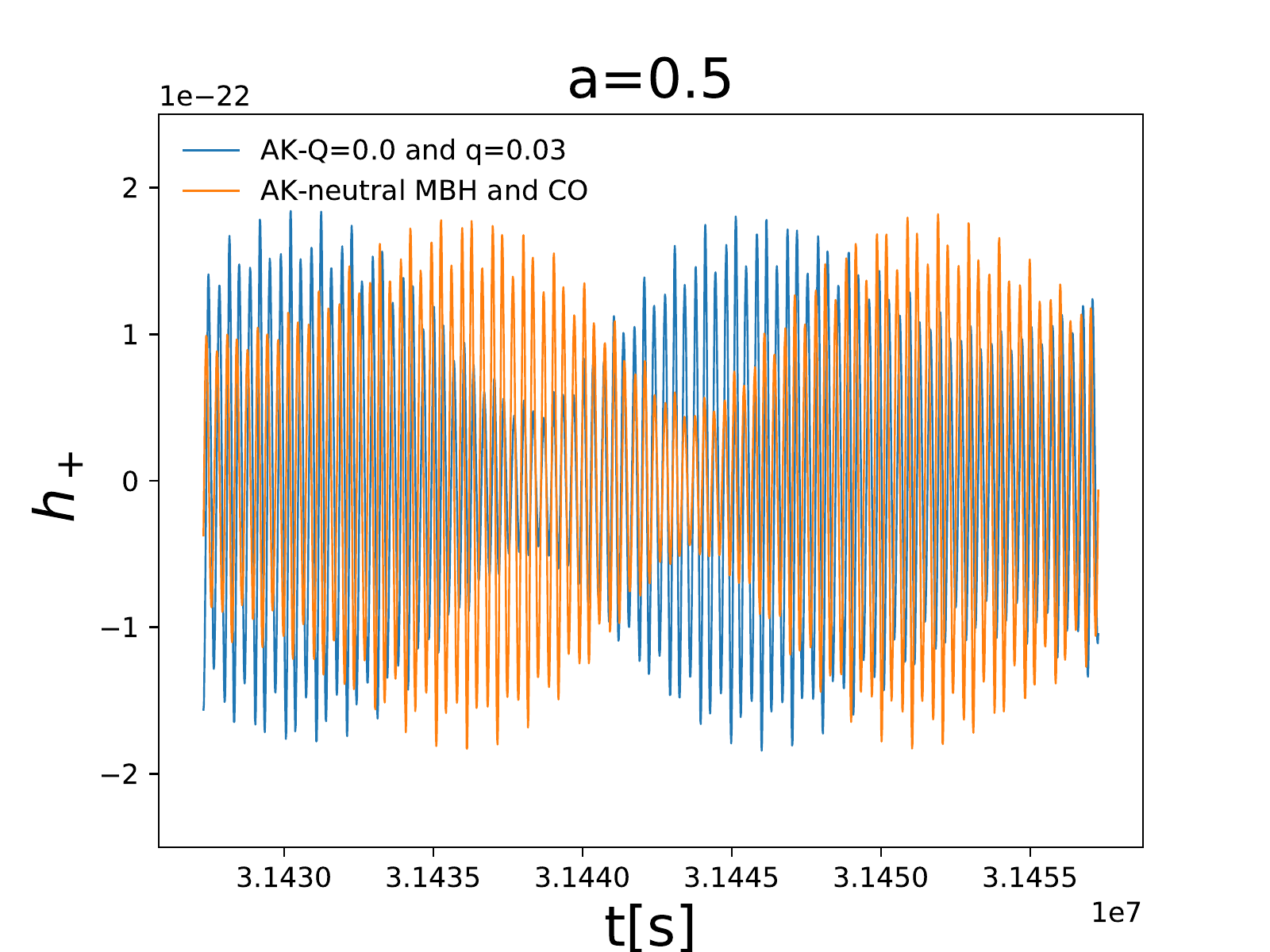}
	\includegraphics[width=0.45\textwidth]{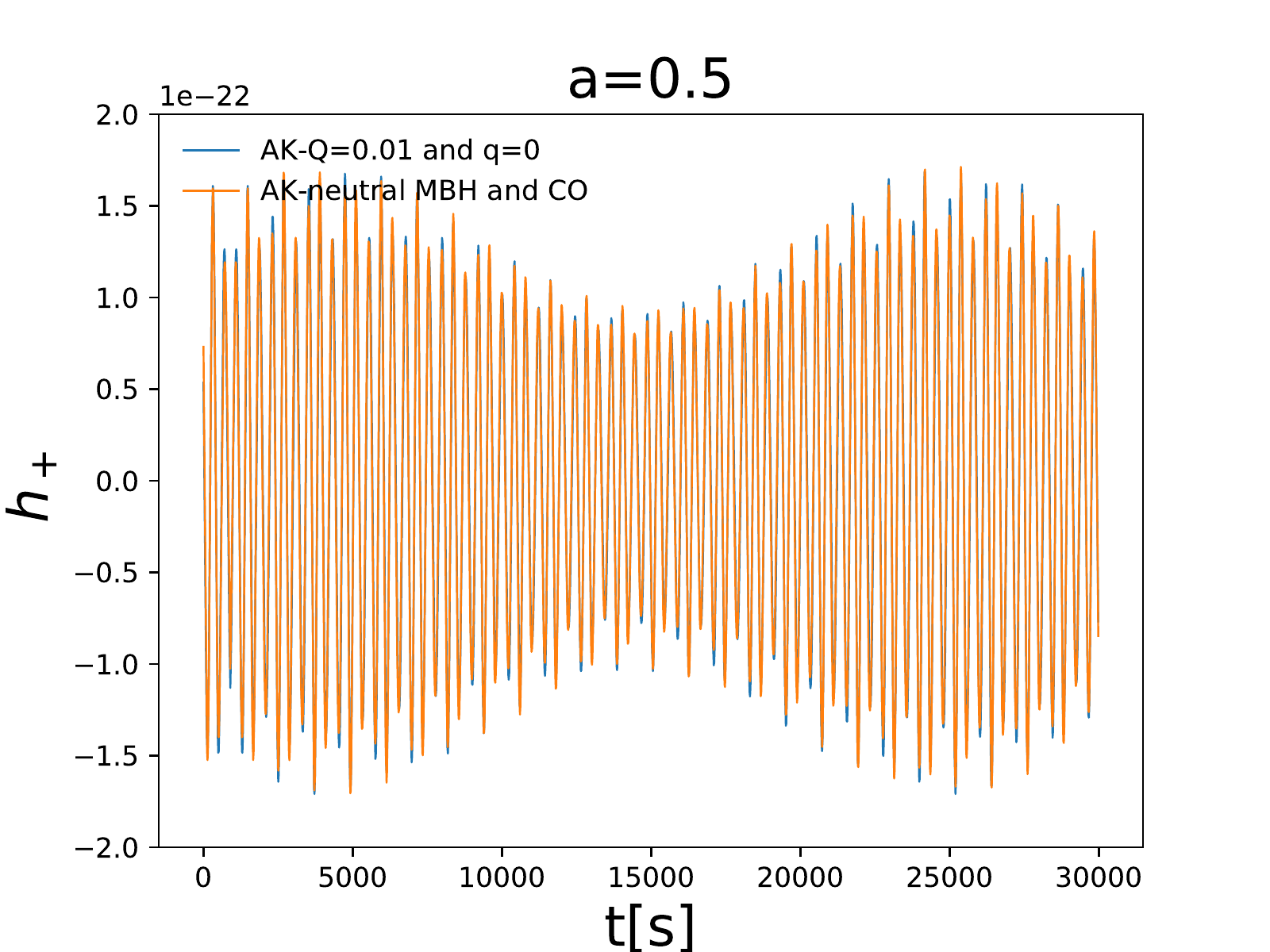}
	\includegraphics[width=0.45\textwidth]{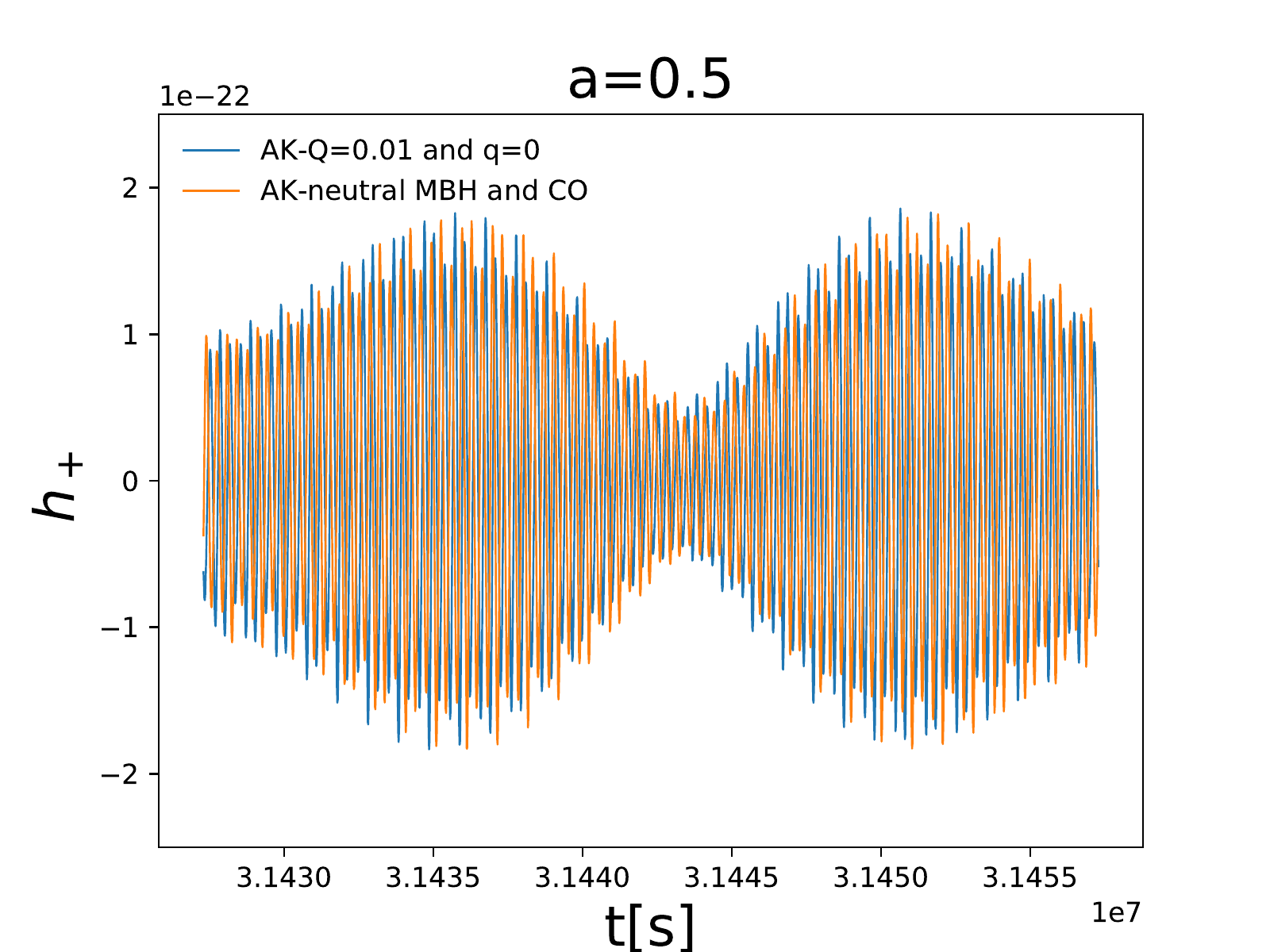}
	\includegraphics[width=0.45\textwidth]{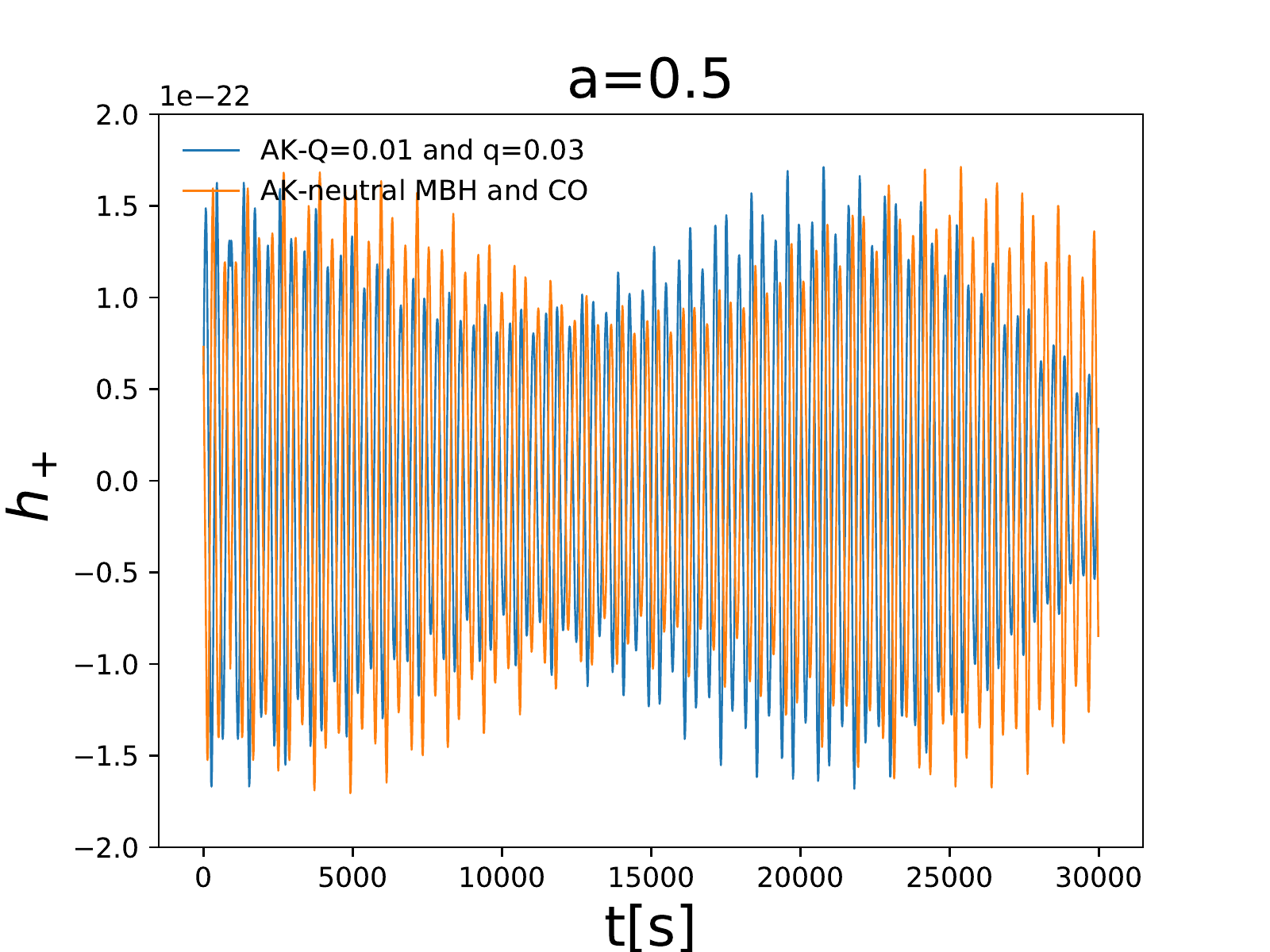}
	\includegraphics[width=0.45\textwidth]{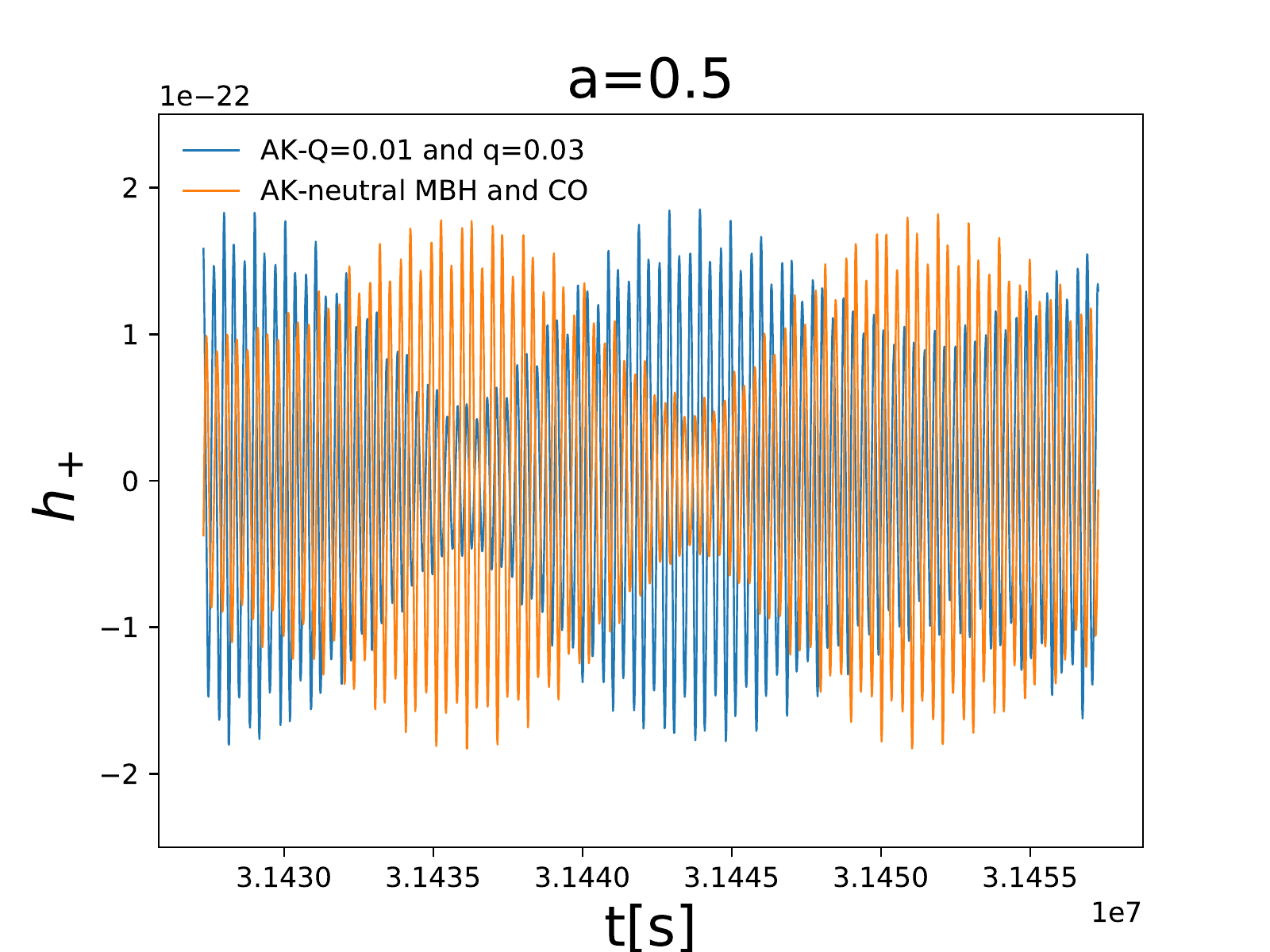}
	\caption{Comparison among plus polarization $h_+$ of AK waveforms from charged EMRI in the case of spin $a=0.5$ for several examples, where the initial frequency is set as $\nu_0=1 \textup{mHz}$.
 Top panels: the CO is charged with $q=0.03$ and the central BH is neutral with $Q=0$.
 Middle panels: the central BH is charged with $Q=0.01$ and the CO is neutral with $q=0$.
 Bottom panels: both the MBH and the CO are charged with $Q=0.01$ and charged CO with $q=0.03$.
 The other parameters are set as Sec.\ref{source:params}.
 The length of the waveform is 1 year, and the left panels represent the waveform for the first 30000 seconds, while the right panels for the last 30000 seconds.
 }\label{waveformCharge:timefrequency:a05}
\end{figure*}

By solving above orbital evolution equations \eqref{evolution} numerically, we can  plot the charged AK waveforms in the time domain.
In Fig.~\ref{waveformCharge:timefrequency:a05}, we show the comparison of the charged AK waveforms with the neutral ones in various cases. Since we focus on the charge parameters in the waveform, we let the other parameters in \eqref{parameters} to be fixed as follows:
$t_0=1$ years, $D=1$ Gpc, $m=10~{\rm M}_\odot$, $M=10^{6}~{\rm M}_\odot$, $e=0.1$,
$\lambda=\pi/3$, $\tilde{\gamma}_0=5\pi/6$,
$\alpha_0=4\pi/5$, $\theta_S=\pi/5$,
$\phi_S=\pi/4$, $\theta_K=2\pi/3$, $\phi_K=3\pi/4$, $\Phi_0=\pi/3$, and $\nu_0=1 \text{mHz}$. Note that for comparison,
when plotting the waveforms we set the
initial radial orbital frequency of the charged and neutral cases to be the same, so here $t_0$ denotes the duration time of the waveforms.

\begin{figure}[th]
	\centering
	\includegraphics[width=0.45\textwidth]{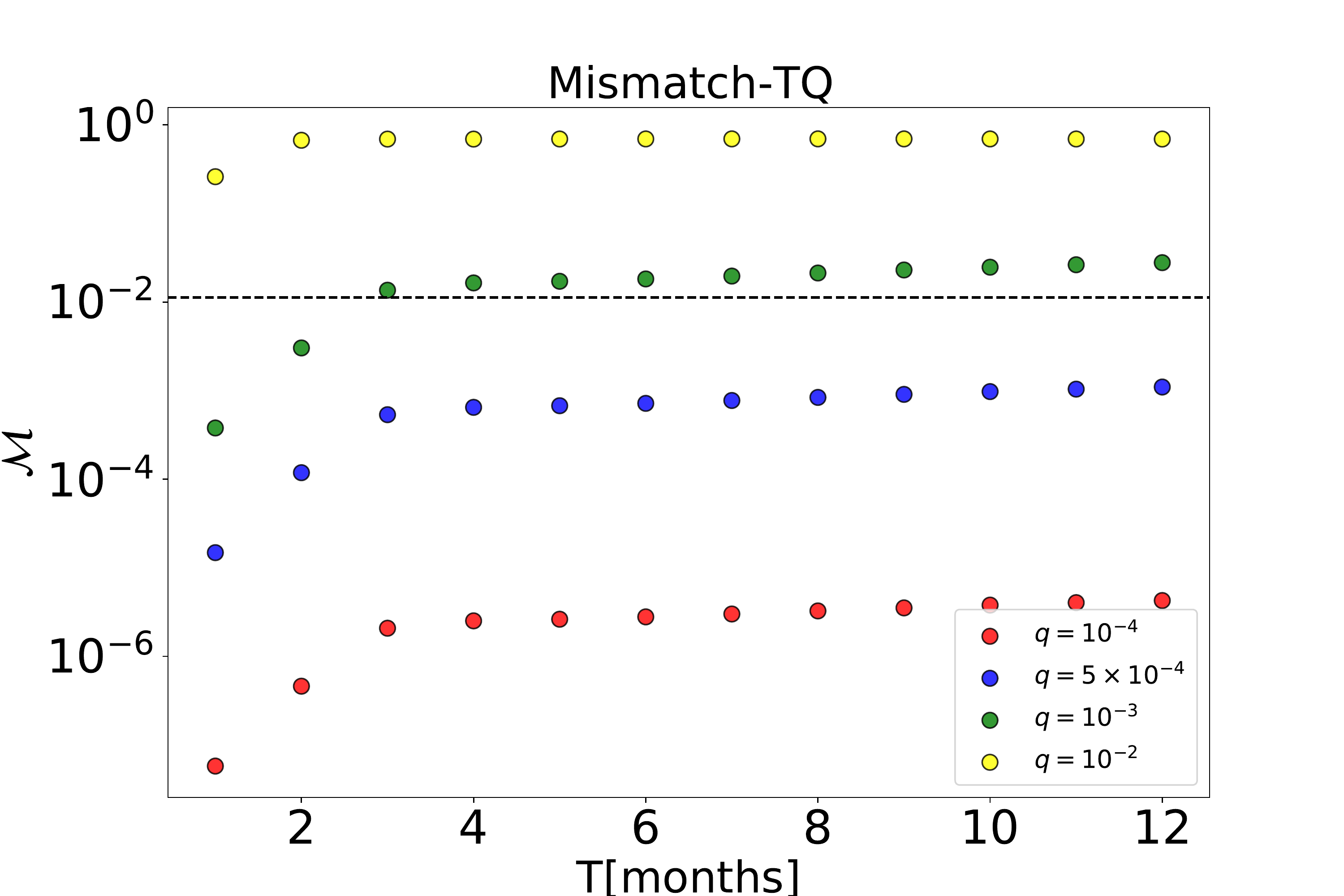}
	\includegraphics[width=0.45\textwidth]{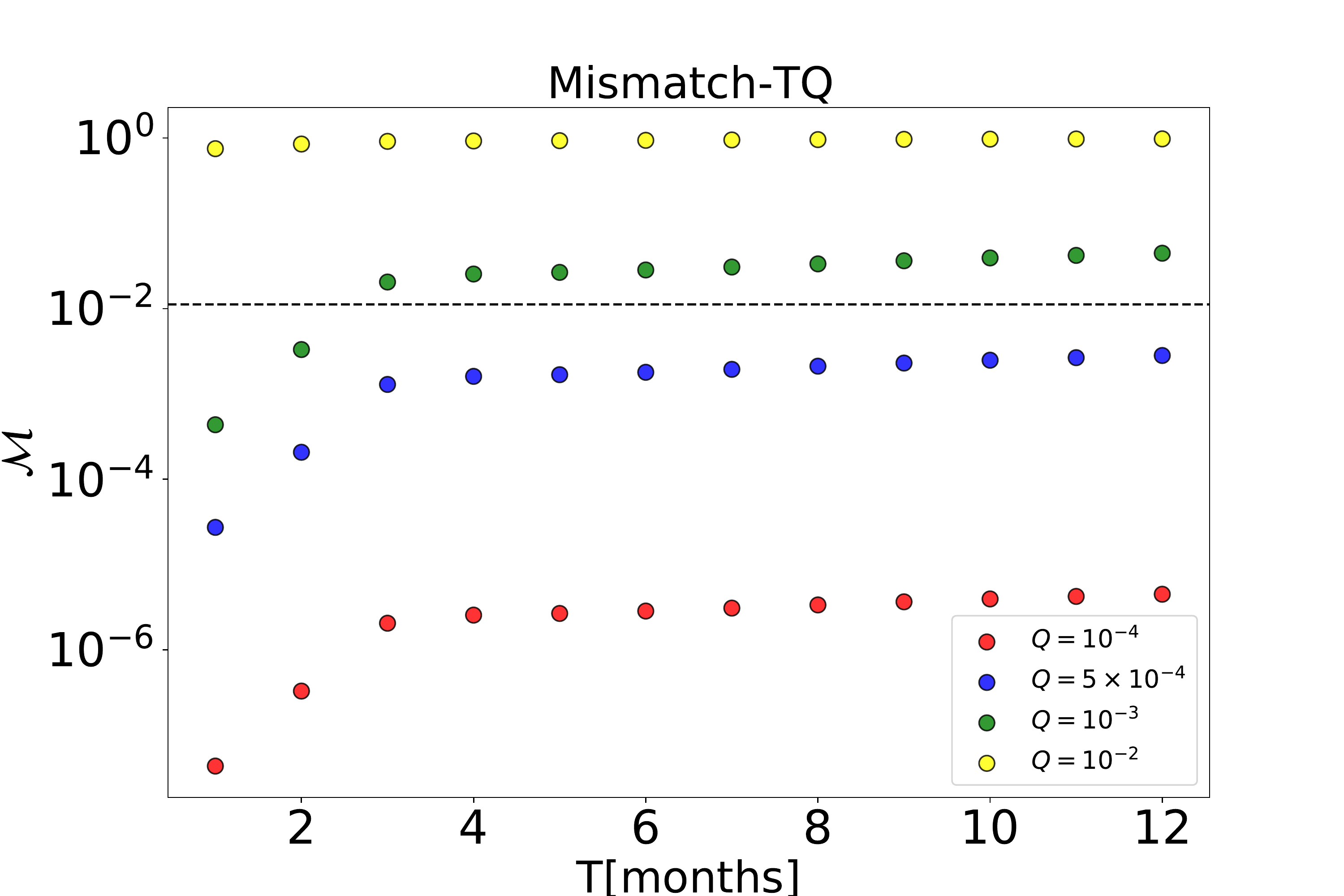}
	\includegraphics[width=0.45\textwidth]{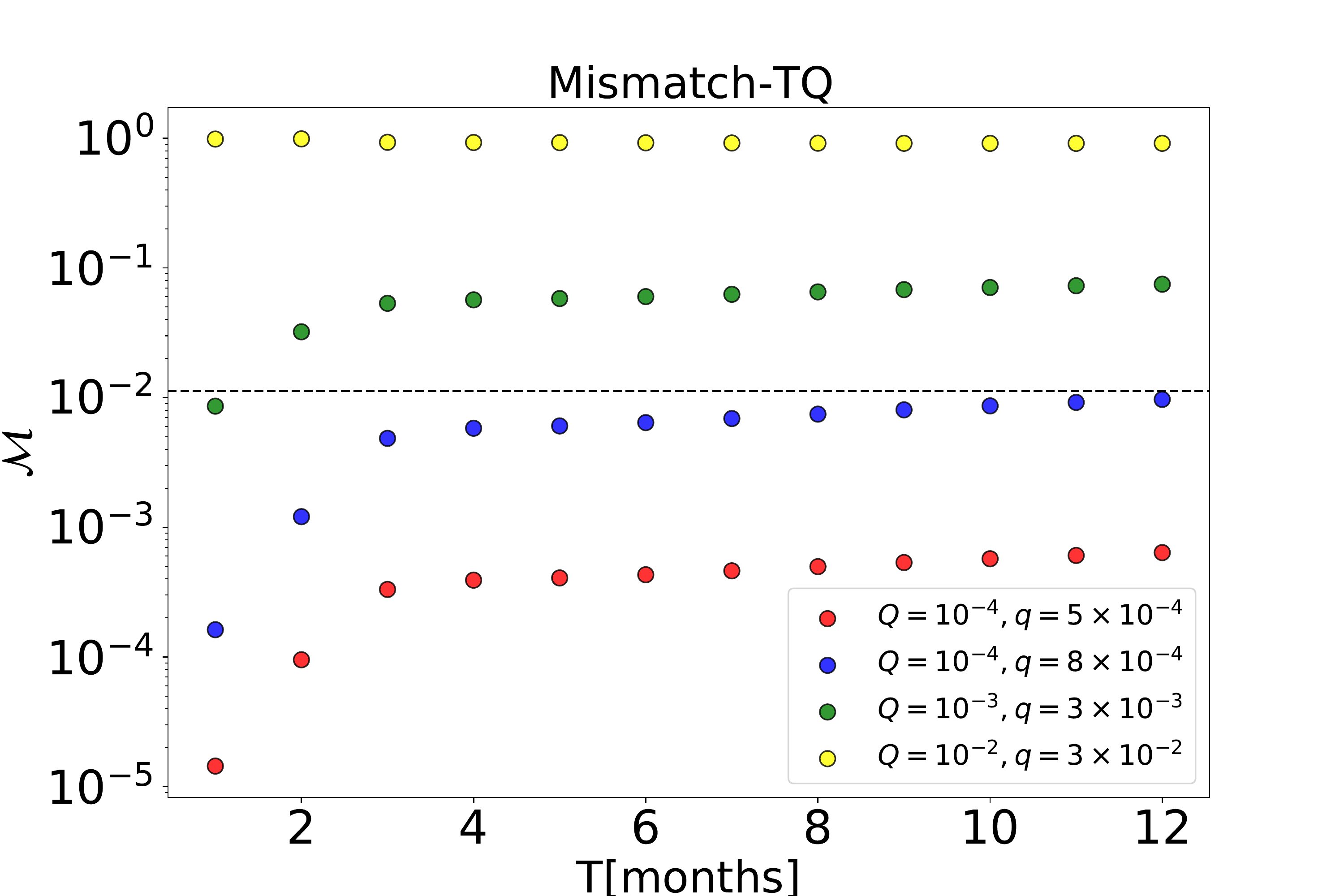}
	\caption{The mismatch $\mathcal{M}$ as a function of observation time for TianQin is plotted,the dashed lines represent the threshold for SNR=20.
	The source parameters are set as $M=10^6M_\odot$, $a=0.5$,  the charge $Q$ of MBH is 0 in the top panel
	and the charge of CO is 0 in the middle panel,
	 the other parameters keep same with the previous configurations in Fig.~\ref{waveformCharge:timefrequency:a05}.
		}\label{mismatch:ObservationTime:TQ}
\end{figure}

From Fig.~\ref{waveformCharge:timefrequency:a05} we can see that  the AK waveforms are significantly affected by the charges of the system.
In this case the initial radial orbital frequencies of two EMRI systems are set to be the same as each other, so the waveforms at $t=0$ should also be identical, and the effects of charges on waveform will grow with the increasing time.
Particularly, when both members of the EMRI are charged, the dephasing occurs very quickly in the first 30000 seconds at the beginning as showed by the panel on the left of the bottom.
If only the CO is charged, and the MBH is neutral, although the waveform is almost the same for the case that both objects are neutral, by a long accumulation of time, the dephasing is still visible from the top panel on the right.
For the case that only the MBH is charged, and the CO is neutral, the dephasing is not very significant even after 1 year as showed by the right panel in the middle, but the mismatch also exceed the threshold as plotted in Fig.~\ref{mismatch:ObservationTime:TQ}. In fact if the MBH carries the same amount of charges as the CO, the dephasing of the former would be more prominent, as we will see below.

To assess the effect of charge on EMRI waveform quantitatively,
we calculate the mismatch for neutral waveform and the charged waveform.
As shown in Fig.~\ref{mismatch:ObservationTime:TQ},  the mismatch is plotted as the function of observation time for TQ,
where the initial radial orbital frequencies for the two  waveforms are set as $1\rm mHz$ and the SNR of the signal is set to be 20.
According to Fig.~\ref{mismatch:ObservationTime:TQ}, the mismatch can exceed the threshold value $\mathcal{M}_{\rm min} = 0.01125$
even though the charges of EMRI system are small to $10^{-3}$.
From the upper two panels of Fig.~\ref{mismatch:ObservationTime:TQ} where only one member carries charge, one can see that
TQ can distinguish the  modified waveform with charge $Q,q\approx10^{-3}$ in three months.
When both the members of the EMRI carry charge $Q=10^{-3}, q=3\times10^{-3}$ as shown in the bottom panel,
the effects of charges on AK waveform would be recognized within two months.
However, if both of the two objects carries the charges less than the level of $10^{-4}$, the missmatch will not exceed the threshold, and thus we cannot distinguish whether the CO and the MBH have charges or not.
Furthermore, we can see that the MBH charge $Q$ yields slightly different influence on the AK waveforms than that of the CO charge $q$,
it is because that the effects of MBH charge and CO charge on the waveform are not equal, the former can also affect the waveform through its deformation on the metric.

\begin{figure}[th]
	\centering
	\includegraphics[width=0.45\textwidth]{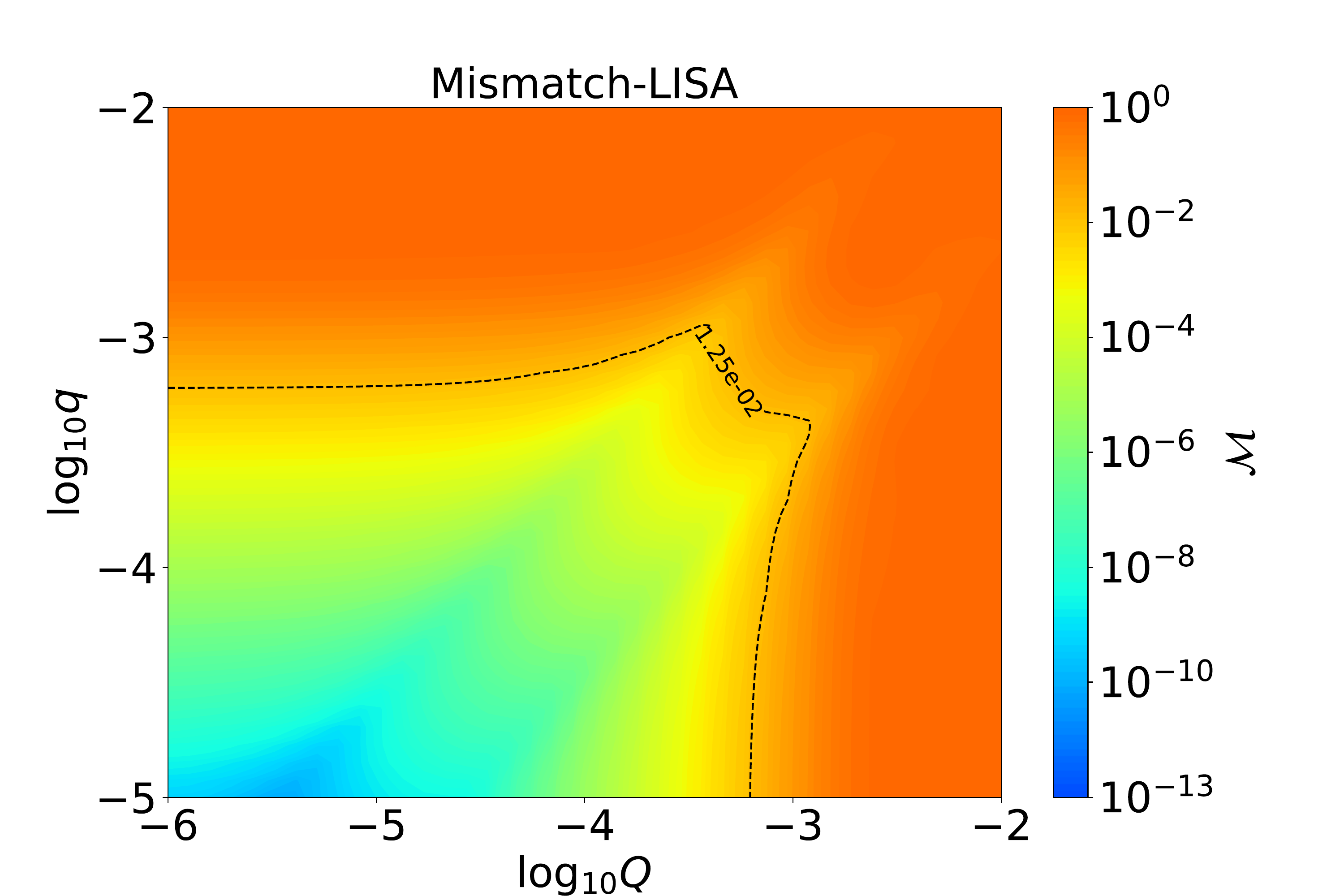}
	\includegraphics[width=0.45\textwidth]{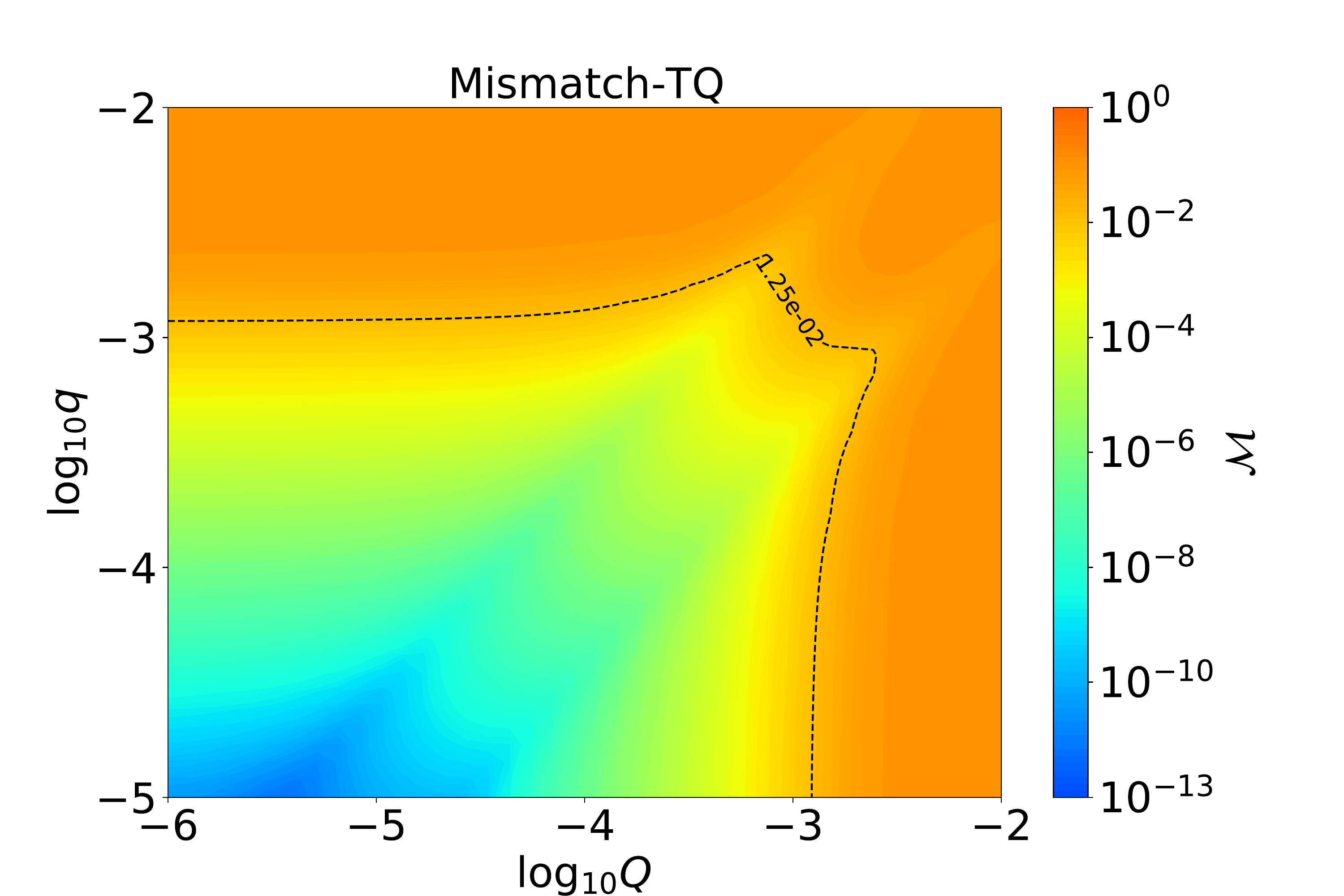}
	\caption{The contour plot of the mismatch $\mathcal{M}$ as a function of $\log_{10} {Q}$ and  $\log_{10} {q}$ with respect to LISA (top panel) and TQ (bottom panel), respectively.
	The black dashed line denotes to the threshold value for SNR=20, where the source parameters are set as $M=10^6M_\odot$, $a=0.4$, and
the other parameters keep same with the previous configurations in Fig.~\ref{waveformCharge:timefrequency:a05}.
		}\label{mismatchContourf}
\end{figure}

To assess the effects of charges on the waveform more clearly,
we plot the mismatch as a function of both $Q$ and $q$ for LISA and TQ respectively
in Fig.~\ref{mismatchContourf}.
As it is shown, the black dotted line represents the contour of mismatch equal to the threshold $\mathcal{M}= 0.01125$,
it indicates that the space-borne GW detectors can distinguish whether the objects in an EMRI is charged if the system is located beyond this curve.
The behaviour of mismatch contour plots for LISA and TQ is almost the same,
the threshold values for charges are about $Q\approx10^{-3}$ and $q\approx10^{-3}$ respectively, and the value for LISA is lower than TQ, since it will have better performance in lower frequency band.

\begin{figure}[th]
	\centering
	\includegraphics[width=0.45\textwidth]{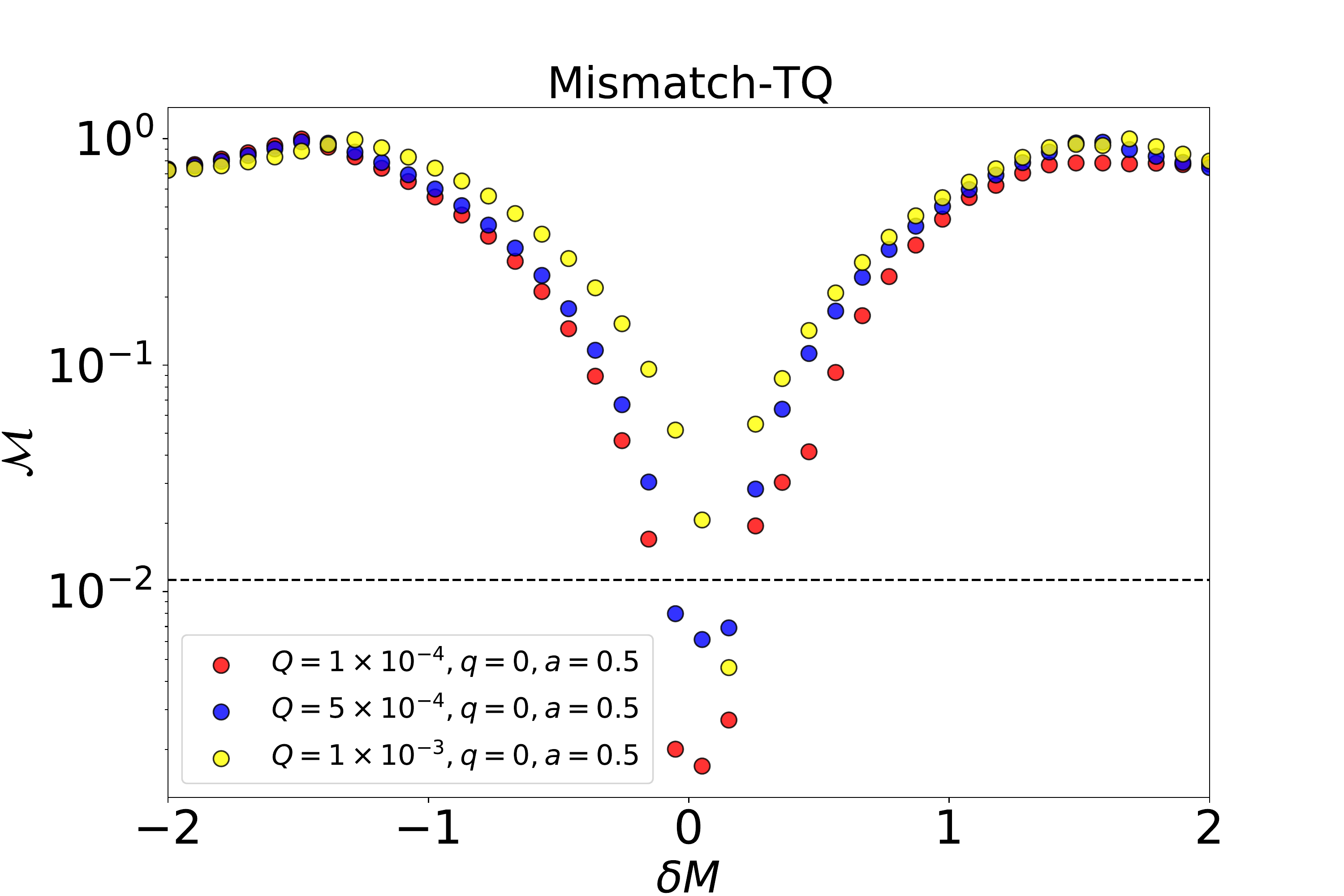}
	\caption{The mismatch $\mathcal{M}$ as a function of  mass variation $\delta M$ of MBH  for TianQin.
	The source parameters are set as $M=(10^6 + \delta M)M_\odot$, $a=0.5$ and $q=0$
	 the other parameters keep same with the previous configurations in Fig.~\ref{waveformCharge:timefrequency:a05}.
		}\label{mismatch:deltaMass:TQLISA}
\end{figure}

To study whether the presence of charges will cause some bias on the parameter estimation even if the mismatch does not exceed the threshold, we calculate the mismatch between the waveforms from a charged EMRI and from a neutral EMRI keeping all the parameters the same as the previous one except the mass of the MBH.
The deviation of $M$ is noted as $\delta M$, we keep this parameter $M$ changing because MBH mass will dominate the orbital frequency and it's evolution for an EMRI system, and the existence of charges will also influence the evolution of the frequency .
In Fig. \ref{mismatch:deltaMass:TQLISA} we plot the mismatch as a function of $\delta M $ for TQ, one can see that for given charges the threshold of the mismatch is satisfied only when $|\delta M |$ is much smaller than $M_\odot$, and it has already exceed the accuracy for the measurement of $M$ which is about $\sim 10^{-6}$ relative to $M$ \cite{Fan:2020zhy}.
Thus, we can conclude that the presence of charges will not affect the parameter estimation precision of the mass of the central BH, if we cannot distinguish whether it is charged or not.

\subsection{Constraint on charges}

In this subsection we perform the parameter estimation for the charges $Q$ and  $q$ using the FIM method.
Note that in this subsection to characterize the effects of the charges on the parameter estimation, the cutoff of the inspiral is chosen to be the ISCO, such that the charged and uncharged waveforms have different cutoff frequencies.
It should be noted that charges indeed influence on the radius of ISCO and parameter estimation is subjected with the cutoff frequency\cite{Zi:2021pdp},
thus it is necessary to assess quantitatively the effects of charges on parameter estimation.
We also choose the luminosity to be 1 Gpc, and do not normalize the SNR, since it will significantly influence the parameter estimation accuracy.

First of all,  by taking  the central values of the charges to be zero, we can study the effects of the mass $M$ and the spin parameter $a$ on the constraints for the charges.  The constraints for $Q$ and $q$ measured by LISA and TQ, are showed in Table \ref{DeltaQtable} and \ref{Deltaqtable} respectively. One can see that in the chosen range for $M$ and  $a$, constraints achievable for $Q$ and $q$ are in the range of $10^{-1}\sim 10^{-5}$. Overall, The capability for TQ and LISA are almost on the same level, while LISA is a little bit better, since the generated GWs are in the lower frequency band. The constraints will be better for an EMRI system with lighter mass and higher spin.

\begin{table*}[!htbp]
	\caption{$\Delta Q$ for different mass $M$ and spin $a$ of the MBH is listed.
	The plain and bold values correspond to the observation of LISA and TQ respectively. Two charge parameters $q$ and $Q$ both are set as zero. The other parameters keep same with the previous configurations in Fig.~\ref{waveformCharge:timefrequency:a05}. }\label{DeltaQtable}
	\begin{center}
		\setlength{\tabcolsep}{5mm}
		\begin{tabular}{|c|c|c|c|c|c|}
			\hline
			\multirow{2}{*}{$a$} & \multicolumn{5}{|c|}
			{MBH mass $\log_{10}(M/M_\odot)$}\\
			\cline{2-6}
			 & $5.0$  &$5.5$ &$6.0$  &$6.5$ & $7.0$ \\
            \hline
			$0.01$ & $4.5\times10^{-3}~~ $ &$7.8\times10^{-3}$ &$2.1\times10^{-1}$
			&$-$    &$-$\\	
            &$\boldsymbol{4.7\times10^{-3}}$ &$\boldsymbol{9.1\times10^{-2}}$
            &$\boldsymbol{3.2\times10^{-1}}$ &$\boldsymbol{-}$
            &$\boldsymbol{-}$        \\		
			\hline
			$0.15$ &$3.2\times10^{-3}$    &$6.8\times10^{-3}$& $4.8\times10^{-2}$
			&$-$  &$-$\\
            &$\boldsymbol{4.2\times10^{-3}}$  &$\boldsymbol{2.4\times10^{-2}}$
            &$\boldsymbol{1.2\times10^{-1}}$  &$\boldsymbol{-}$
            &$\boldsymbol{-}$ \\
			\hline
			$0.25$ &$2.5\times10^{-3}$ &$4.6\times10^{-3} $  &$4.7\times10^{-2}$
			&$2.4\times10^{-1}$ & $-$\\
		    &$\boldsymbol{3.9\times10^{-3}}$  &$\boldsymbol{7.8\times10^{-3}}$
            &$\boldsymbol{8.8\times10^{-2}}$ &$\boldsymbol{7.7\times10^{-1}}$
            &$\boldsymbol{-}$      \\
			\hline
			$0.35 $&$ 2.3\times10^{-3} $&$4.3\times10^{-3} $&$2.9\times10^{-2}$
            &$1.7\times10^{-1}$&$-$\\			
		    &$\boldsymbol{3.8\times10^{-3}}$ &$\boldsymbol{6.2\times10^{-3}}$
            &$\boldsymbol{7.2\times10^{-2}}$   &$\boldsymbol{3.4\times10^{-1}}$
            &$\boldsymbol{-}$     \\
			\hline
			$0.45$&$2.2\times10^{-3} $&$3.1\times10^{-3} $&$1.9\times10^{-2}$
			&$6.8\times10^{-2}$     &$-$  \\
		    &$\boldsymbol{3.6\times10^{-3}}$  &$\boldsymbol{4.3\times10^{-3}}$
            &$\boldsymbol{6.1\times10^{-2}}$  &$\boldsymbol{ 2.1\times10^{-1}}$
            &$\boldsymbol{-}$   \\
			\hline
			$0.55 $ &$ 2.1\times10^{-3} $&$2.6\times10^{-3}$ &$1.4\times10^{-2}$
            &$5.3\times10^{-2}$ &$-${\tiny }\\
		    &$\boldsymbol{3.4\times10^{-3}}$  &$\boldsymbol{2.8\times10^{-3}}$
            &$\boldsymbol{2.5\times10^{-2}}$ &$\boldsymbol{1.4\times10^{-1}}$
            &$\boldsymbol{-}$ \\
			\hline
			$0.65$  &$1.8\times10^{-3} $&$2.3\times10^{-3} $&$9.5\times10^{-3}$
			&$4.3\times10^{-2}$  &$9.4\times10^{-1}$\\
            &$\boldsymbol{3.2\times10^{-3}}$  &$\boldsymbol{2.5\times10^{-3}}$
            &$\boldsymbol{1.8\times10^{-2}}$  &$\boldsymbol{9.2\times10^{-2}}$
            &$\boldsymbol{8.6\times10^{-1}}$\\
		    \hline
			$0.75$  &$ 1.5\times10^{-3} $&$2.3\times10^{-3} $&$5.8\times10^{-3}$
			&$4.3\times10^{-1}$ &$3.7\times10^{-1}$ \\
            &$\boldsymbol{1.7\times10^{-3}}$  &$\boldsymbol{2.4\times10^{-2}}$
            &$\boldsymbol{6.1\times10^{-3}}$  &$\boldsymbol{3.4\times10^{-2}}$
            &$\boldsymbol{4.5\times10^{-2}}$\\
			\hline
			$0.85$&$ 9.7\times10^{-4} $&$1.6\times10^{-3} $&$1.5\times10^{-3}$
			&$1.4\times10^{-2}$ &$1.1\times10^{-1}$\\
		    &$\boldsymbol{1.6\times10^{-3}}$  &$\boldsymbol{1.8\times10^{-3}}$
            &$\boldsymbol{3.7\times10^{-3}}$  &$\boldsymbol{7.7\times10^{-3}}$
            &$\boldsymbol{3.3\times10^{-2}}$ \\
			\hline
			$0.95$&$ 1.6\times10^{-4} $&$8.8\times10^{-4} $&$1.3\times10^{-3} $
			&$6.5\times10^{-3}$&$5.1\times10^{-2}$\\
		    &$\boldsymbol{3.5\times10^{-4}}$   &$\boldsymbol{1.3\times10^{-3}}$
            &$\boldsymbol{2.7\times10^{-3}}$ &$\boldsymbol{2.3\times10^{-3}}$
            &$\boldsymbol{2.1\times10^{-2}}$\\
			\hline
		\end{tabular}
	\end{center}
\end{table*}

\begin{table*}[!htbp]
	\caption{$\Delta q$ for different mass $M$ and spin $a$ of the MBH is listed.
	The plain and bold values correspond to the observation of LISA and TQ respectively.
	Two charge parameters $q$ and $Q$ both are set as zero.
 The other parameters keep same with the previous configurations in Fig.~\ref{waveformCharge:timefrequency:a05}.}\label{Deltaqtable}
	\begin{center}
		\setlength{\tabcolsep}{5mm}
		\begin{tabular}{|c|c|c|c|c|c|}
			\hline
			\multirow{2}{*}{$a$}& \multicolumn{5}{|c|}
			{MBH mass $\log_{10}(M/M_\odot)$}\\
			\cline{2-6}
			& $5.0$ &$5.5$ &$6.0$ &$6.5$ & $7.0$ \\
            \hline
			$0.01$ & $1.3\times10^{-3}~~ $ &$1.4\times10^{-3}$ &$4.6\times10^{-2}$
			&$8.3\times10^{-1}$  &$-$\\	
            &$\boldsymbol{1.9\times10^{-3}}$  &$\boldsymbol{7.1\times10^{-2}}$
            &$\boldsymbol{1.7\times10^{-1}}$   &$\boldsymbol{8.8\times10^{-1}}$
            &$\boldsymbol{-}$
             \\		
			\hline
			$0.15$ &$1.1\times10^{-3}$ &$1.3\times10^{-3}$& $9.4\times10^{-3}$
			&$5.1\times10^{-1}$  &$-$\\
            &$\boldsymbol{8.2\times10^{-3}}$  &$\boldsymbol{1.7\times10^{-2}}$
            &$\boldsymbol{1.1\times10^{-1}}$   &$\boldsymbol{7.5\times10^{-1}}$
            &$\boldsymbol{-}$
             \\
			\hline
			$0.25$ &$ 7.4\times10^{-4}$  &$8.4\times10^{-4} $  &$9.2\times10^{-3} $
			&$3.7\times10^{-1}$     & $-$\\
		    &$\boldsymbol{6.2\times10^{-3}}$  &$\boldsymbol{5.6\times10^{-3}}$
            &$\boldsymbol{6.4\times10^{-2}}$   &$\boldsymbol{4.7\times10^{-1}}$
            &$\boldsymbol{-}$
            \\
			\hline
			$0.35 $&$5.6\times10^{-4} $  &$7.9\times10^{-4} $  &$5.8\times10^{-3}$
             &$2.5\times10^{-1}$    &$-$\\			
		    &$\boldsymbol{5.6\times10^{-3}}$  &$\boldsymbol{4.9\times10^{-3}}$
            &$\boldsymbol{5.2\times10^{-2}}$   &$\boldsymbol{3.4\times10^{-1}}$
            &$\boldsymbol{-}$
             \\
			\hline
			$0.45$&$5.1\times10^{-4} $  &$ 6.5\times10^{-4}$ &$ 3.6\times10^{-3}$
			&$1.7\times10^{-1} $&$-$    \\
		    &$\boldsymbol{3.5\times10^{-3}}$   &$\boldsymbol{4.8\times10^{-3}}$
            &$\boldsymbol{4.6\times10^{-2}}$   &$\boldsymbol{2.5\times10^{-1}}$
            &$\boldsymbol{-}$
            \\
			\hline
			$0.55 $ &$4.7\times10^{-4} $ &$ 6.3\times10^{-4}$ &$2.6\times10^{-3}$
            &$1.2\times10^{-1}$   &$-${\tiny} \\
		    &$\boldsymbol{2.6\times10^{-3}}$   &$\boldsymbol{ 1.1\times10^{-3}}$
            &$\boldsymbol{2.5\times10^{-2}}$   &$\boldsymbol{1.4\times10^{-1}}$
            &$\boldsymbol{-}$
            \\
			\hline
			$0.65$  &$3.5\times10^{-4} $ &$5.8\times10^{-4} $ &$1.8\times10^{-3}$
			&$6.2\times10^{-2}$   &$7.9\times10^{-1}$
             \\
            &$\boldsymbol{1.9\times10^{-3}}$  &$\boldsymbol{ 6.2\times10^{-4}}$
            &$\boldsymbol{1.7\times10^{-2}}$    &$\boldsymbol{6.8\times10^{-2}}$
            &$\boldsymbol{5.4\times10^{-1}}$
              \\
		    \hline
			$0.75$  &$1.6 \times10^{-4}  $&$4.2\times10^{-4} $&$1.1\times10^{-3}$
			&$2.8\times10^{-2}$ &$5.1\times10^{-1}$              \\
            &$\boldsymbol{2.4\times10^{-4}}$   &$\boldsymbol{4.4\times10^{-4}}$
            &$\boldsymbol{1.2\times10^{-2}}$   &$\boldsymbol{1.8\times10^{-2}}$
            &$\boldsymbol{2.2\times10^{-1}}$           \\
			\hline
			$0.85$&$ 1.1 \times10^{-4} $&$3.5\times10^{-4}$  &$4.7\times10^{-4}$
			&$1.1\times10^{-2}$ &$9.3\times10^{-2}$\\
		    &$\boldsymbol{1.9\times10^{-4}}$  &$\boldsymbol{4.1\times10^{-4}}$
            &$\boldsymbol{4.8\times10^{-3}}$    &$\boldsymbol{1.5\times10^{-2}}$
            &$\boldsymbol{6.4\times10^{-2}}$     \\
			\hline
			$0.95$&$ 5.6\times10^{-5} $&$2.8\times10^{-4} $ &$4.5\times10^{-4} $
			&$8.5\times10^{-3}$  &$8.3\times10^{-2}$\\
			&$\boldsymbol{1.1\times10^{-4}}$  &$\boldsymbol{3.2\times10^{-4}}$
            &$\boldsymbol{2.3\times10^{-3}}$       &$\boldsymbol{8.7\times10^{-2}}$
            &$\boldsymbol{6.5\times10^{-2}}$         \\
			\hline
		\end{tabular}
	\end{center}
\end{table*}

\begin{figure*}[th]
\centering
\includegraphics[width=0.45\textwidth]{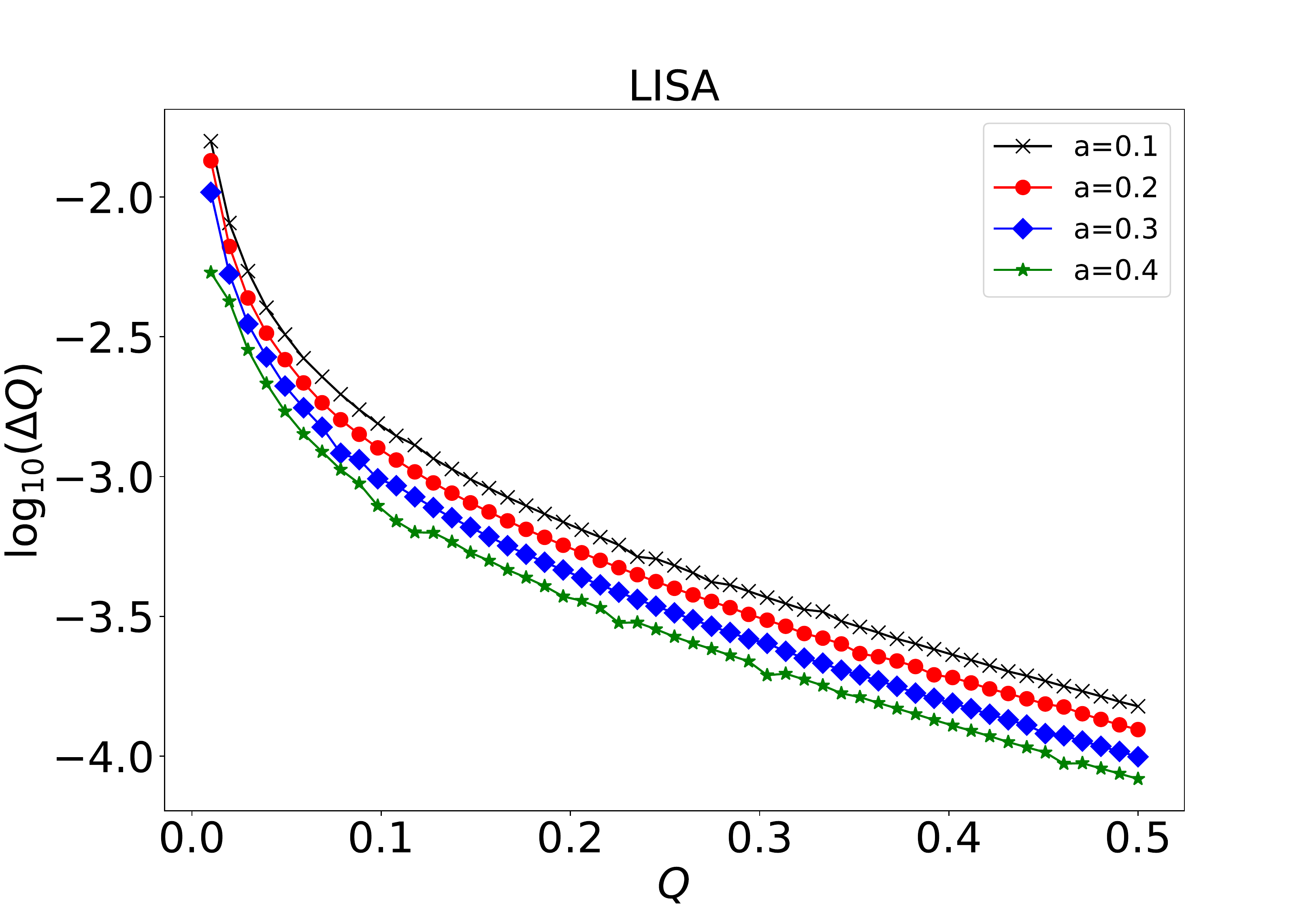}
\includegraphics[width=0.45\textwidth]{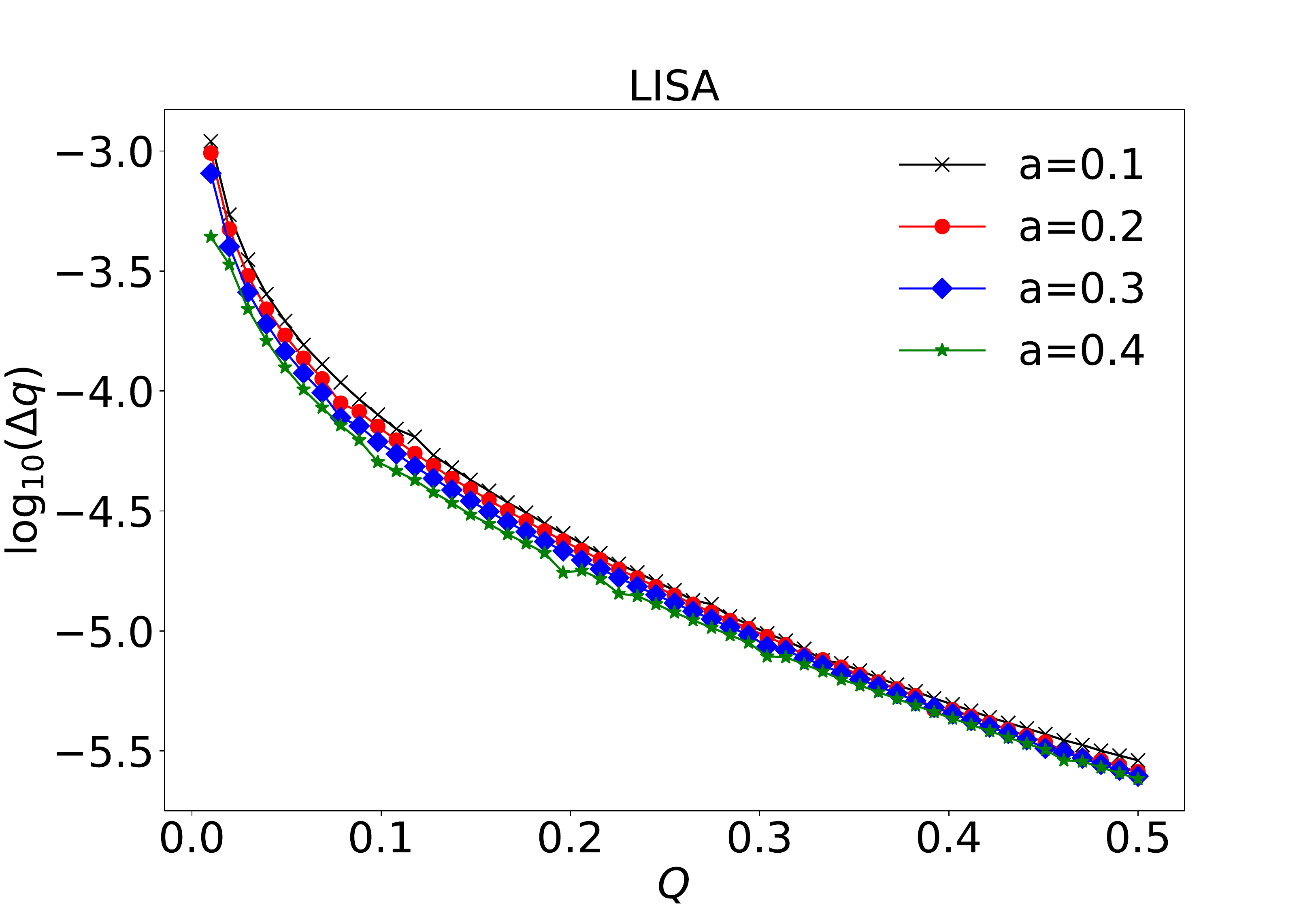}
\includegraphics[width=0.45\textwidth]{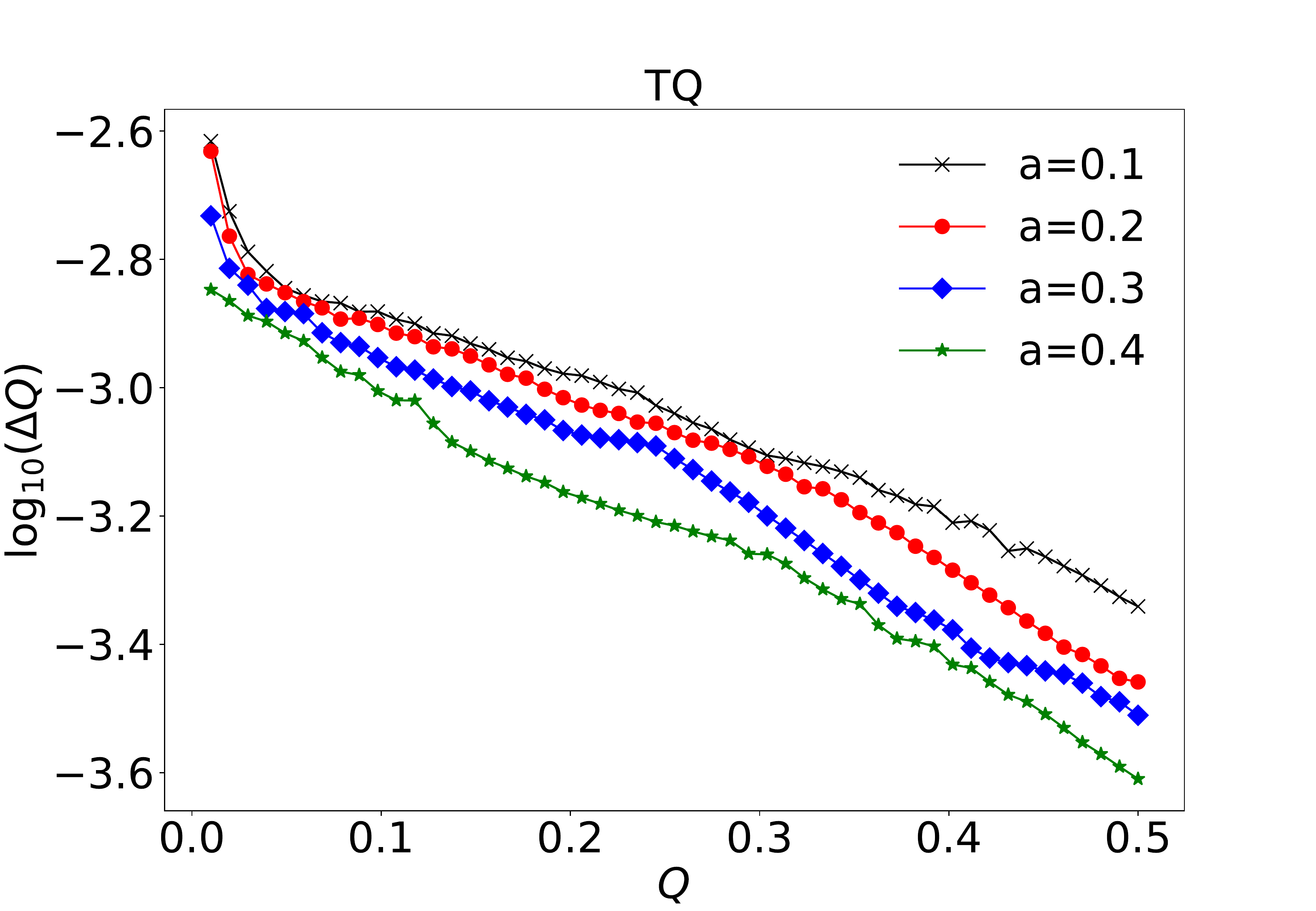}	
\includegraphics[width=0.45\textwidth]{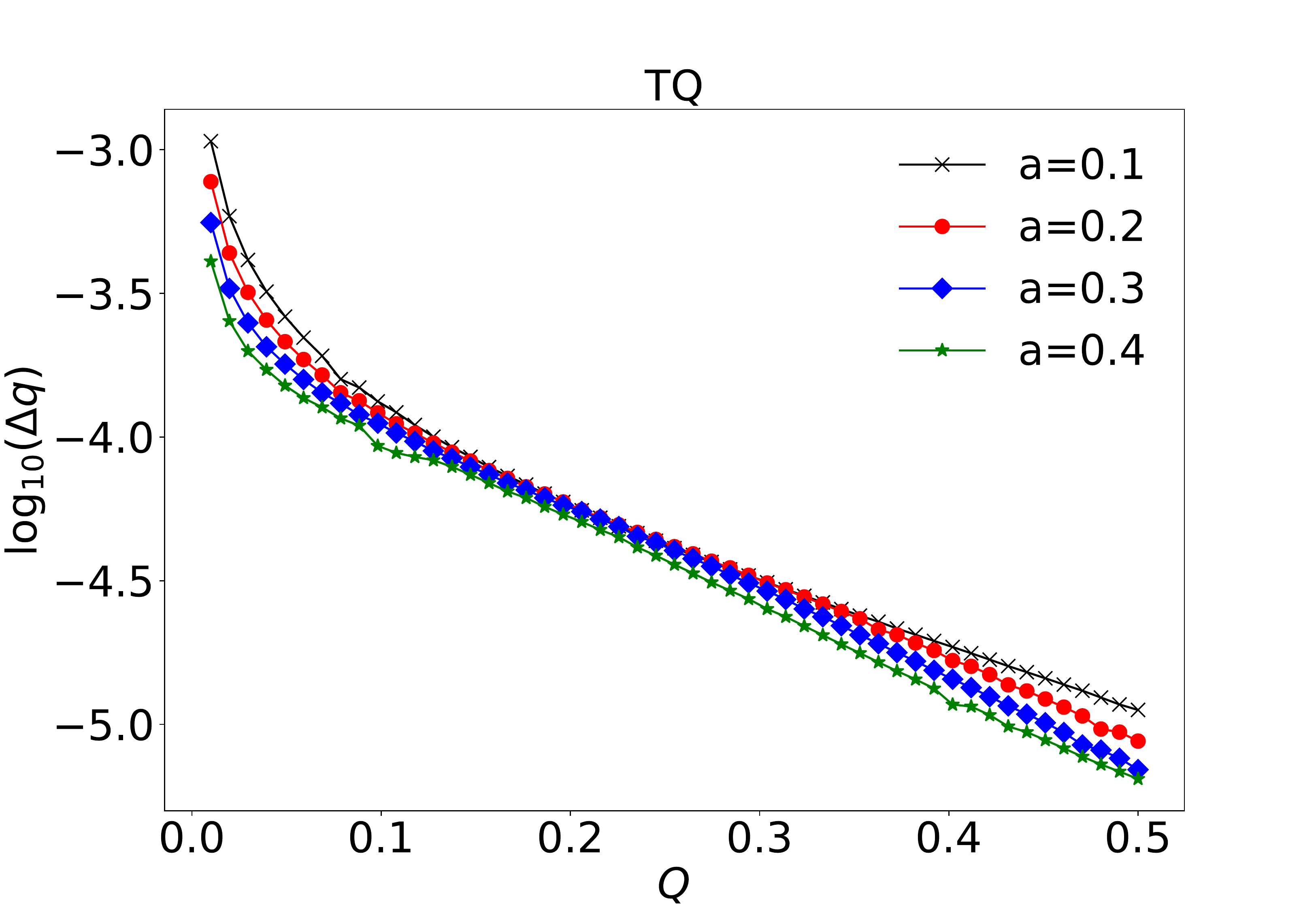}
\caption{PE accuracy for charge parameter, $\log_{10}(\Delta Q)$ (the left panels) and $\log_{10}(\Delta q)$ (the right panels), as a function of MBH charge $Q$ for LISA (the top panels) and TQ (the bottom panels), respectively. 	The charge of CO is set as  $q=0$.	
The other parameters keep same with the previous configurations in Fig.~\ref{waveformCharge:timefrequency:a05}.
}\label{deltaQq}
\end{figure*}

\begin{figure*}[th]
\centering
\includegraphics[width=0.45\textwidth]{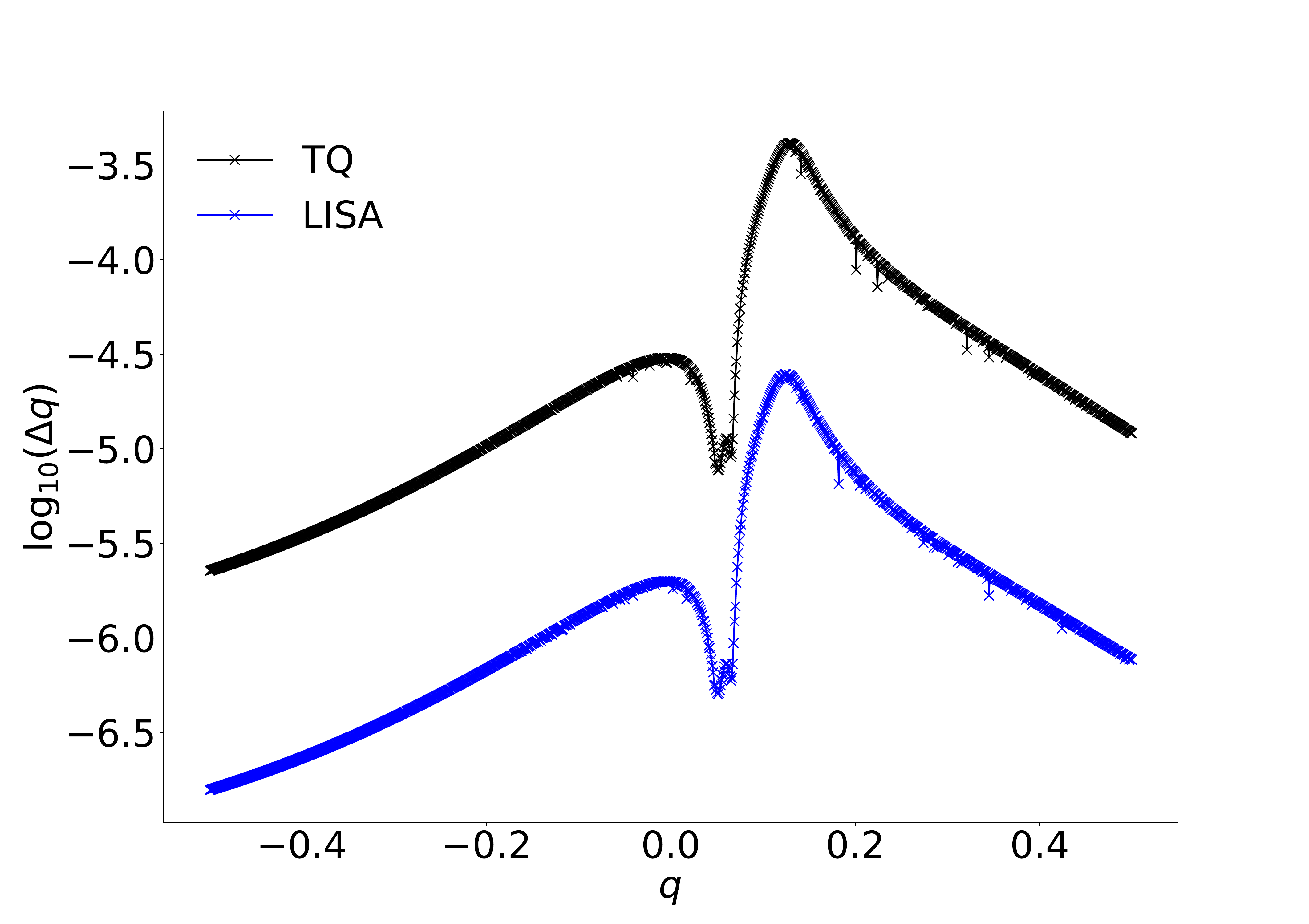}
\includegraphics[width=0.45\textwidth]{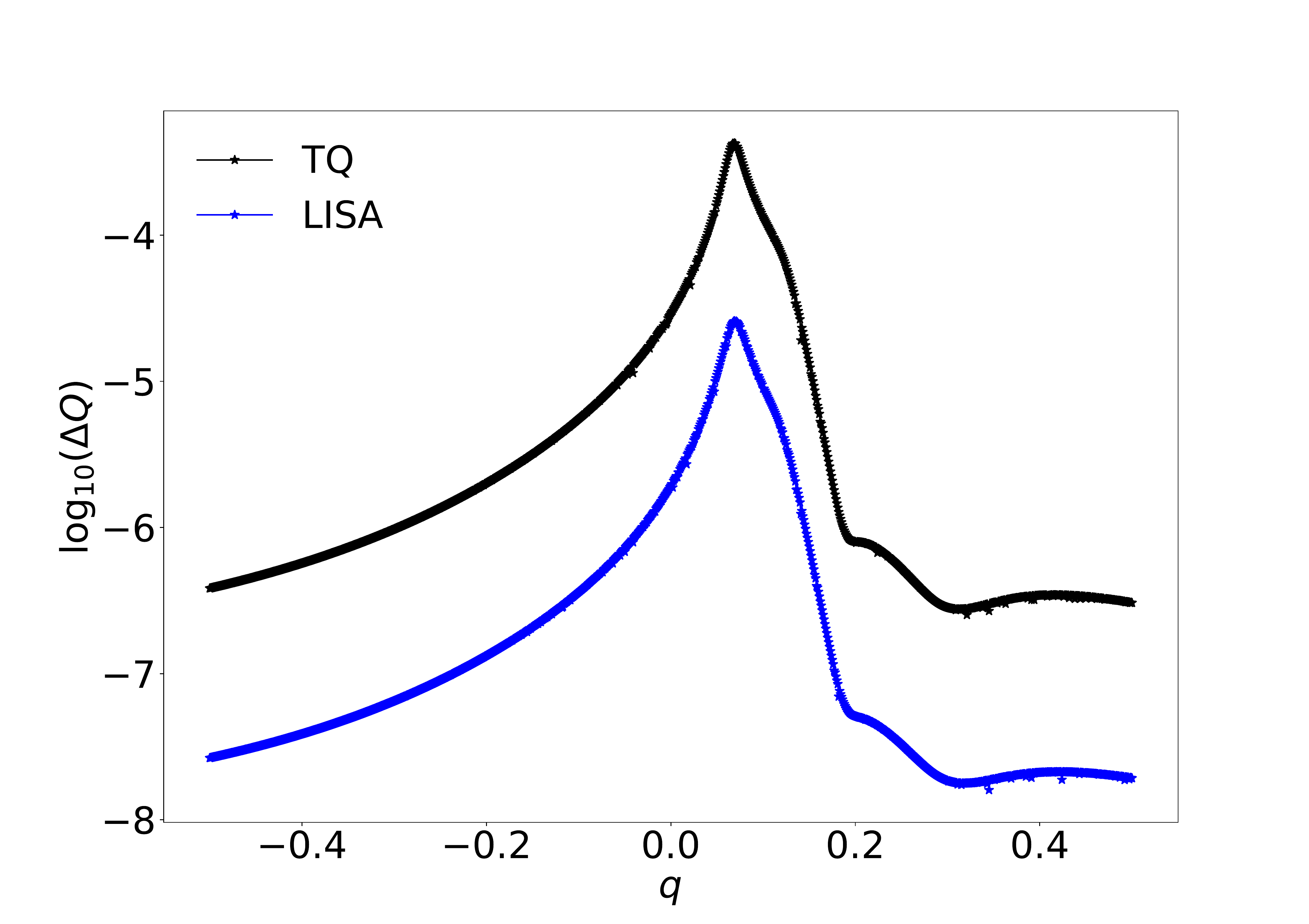}
\caption{PE accuracy for charge parameter, $\log_{10}(\Delta q)$ (the left panel) and $\log_{10}(\Delta Q)$ (the right panel), as a function of CO charge $q$ for LISA (the red circle) and TQ (the black cross) , respectively.
The charge and spin  of MBH is set as  $Q=0.1$ and $a=0.6$, respectively.
The other parameters keep same with the previous configurations in Fig.~\ref{waveformCharge:timefrequency:a05}.	
}\label{deltaQq:q:isco}
\end{figure*}

We then study the effects of the charges on the parameter estimation precision. From Fig.~\ref{deltaQq} we can find that the relative errors for both $Q$ and $q$ decrease with the MBH charge $Q$. This can be explained by the fact that the ISCO radius $r_{\rm ISCO}$ of the charged CO  decreases with the charge of the KN BH, as shown in bottom pane of Fig. \ref{ISCOMBH}.
Smaller $r_{\rm ISCO}$ means the CO will orbiting more cycles around the KN MBH,
then such EMRI event  will have higher SNR.
Similar phenomenon has been found for the strong effect on the parameter estimation from the choices of the waveform cutoff~\cite{Babak:2017tow,Zi:2021pdp}. Besides, we can see that the larger the MBH spin the higher the estimation precision the charges of the waveforms can have. The effects of the CO charge $q$ on  the relative errors for both $Q$ and $q$ are more complicated. As shown in Fig. \ref{deltaQq:q:isco}, we can see that the parameter estimation overall increase when $|q|$ is large, which could be explained by the enhancement of the electric force between the CO and the MBH in this case. However, the unsmooth behavior of $\log(\Delta q)$ at $0<q<0.2$ is unclear.

\section{Conclusion}\label{summary}
In this paper we derived the charged version of the AK waveform by considering the spiral of a charged stellar-mass compact object into a charged massive black hole. The latter is described by the Kerr-Newman metric. From the equations of motion of the charged CO in the KN spacetime, we computed the three fundamental frequencies in the weak-field regime, from which the equations describing the evolution of the perihelion precession and the orbital plane precession were obtained. Moreover, the evolution equations of the radial orbital frequency and the eccentricity were derived from  the energy flux and the angular momentum flux due to the gravitational
radiation and electromagnetic radiation. Combine these leading order corrected
equations with those higher-order PN equations in the original AK model, the complete
orbital evolution equations were obtained.

We found that the correction of charge on AK waveform is evidently different from the original AK waveform,
as long as the EMRI system carry a tiny amount of charges. This is supported  quantitatively  by  calculating the mismatch of the two different AK waveforms with respect to TianQin and LISA. We then performed the parameter estimation precision for the charges $Q$ and $q$ and found that space borne detectors can measure  them with accuracy to the level of $10^{-5}$ under suitable scenarios. This is almost the level of the upper limit caused by different neutralized mechanisms \cite{Barausse:2014tra}, and far beyond the level if some charged mechanisms exist. Moreover, we studied the effects of $Q$ and $q$ on the  parameter estimation precision of themselves. We found that the effects on the parameter estimation precision from $Q$ are almost dominated by the changing of ISCO, and for the parameter estimation precision from $q$, the behavior at large $q$ may due to the
enhancement of the electric force between the CO and
the MBH.

The AK model employed in this work is known to be insufficiently accurate in the strong-field regime. Instead, the numerical kludge (NK) model \cite{Babak:2006uv} is  more accurate but with a slightly expensive computational cost. Therefore, it would be interesting to consider the charged version of the NK model, where the trajectory of CO is obtained by solving the equations of motion strictly and the GW waveform is still calculated
with the  quadrupole-octupole  formula. The last piece of the NK model is using semi-analytic fits to strong-field radiation emission to describe inspiral, which seems not easy for the charged EMRI system. This is because the strong-field radiation emission is governed by the Teuskolsky equations and the counterpart in the charged case is a set of coupled partial differential equations, which is much more harder to handle.

\section*{Acknowledgments}
We are grateful to Prof. W. Schmidt for his kind help on the calculation of the fundamental frequencies and Shuai Liu for his kind help on the calculation of the CDF of logarithm  mismatch ratio. This work is supported by The National Key R\&D Program
of China (Grant No. 2021YFC2203002). This work has been supported by Guangdong Major Project of Basic and Applied Basic Research (Grant No. 2019B030302001), NSFC (Grant NO. 11805286.) P. C. L. is also funded by China Postdoctoral Science
Foundation Grant No. 2020M670010 and the startup
funding of South China University of Technology (Grant No.
D6222420). This project is supported by MOE Key Laboratory of TianQin Project, Sun Yat-sen University.

\end{document}